\documentclass{optica-article}

\journal{opticajournal} 

\articletype{Research Article}

\usepackage{lineno}
\usepackage{amsmath}
\usepackage{mathtools}
\usepackage{booktabs}

\begin{document}

\title{Quantum Circuit Mapping for Universal and Scalable Computing in MZI-based Integrated Photonics}

\author{Yong Kwon,\authormark{1} Alessio Baldazzi,\authormark{2} Lorenzo Pavesi,\authormark{2} and Byung-Soo Choi\authormark{3,*}}

\address{\authormark{1} Department of Physics, Pukyong National University, Korea\\
\authormark{2} Department of Physics, University of Trento, Italy\\
\authormark{3} Department of Scientific Computing, Pukyong National University, Korea}

\email{\authormark{*}bschoi@pknu.ac.kr} 


\begin{abstract*}
Linear optical quantum computing (LOQC) offers a quantum computation paradigm based on well-established and robust technology and flexible environmental conditions following DiVincenzo's criteria.
Within this framework, integrated photonics can be utilized to achieve gate-based quantum computing, defining qubits by path-encoding, quantum gates through the use of Mach-Zehnder interferometers (MZIs) as fundamental building blocks, and measurements through single-photon detectors.
In particular, universal two-qubit gates can be achieved by suitable structures of MZIs together with post-selection or heralding. 
The most resource-efficient choice is given by the post-selected CZ gate.
However, this implementation is characterized by a design which has a non-regular structure and cannot be cascaded.
This limits the implementation of large-scale LOQC.
Starting from these issues, we suggest an approach to move toward a universal and scalable LOQC on the integrated photonic platform.
First of all, choosing the post-selected CZ as universal two-qubit gate, we extend the path-encoded dual-rail qubit to a triplet of waveguides, composed of an auxiliary waveguide and the pair of waveguides corresponding to the qubit basis states.
Additionally, we introduce a swap photonic network that maps the regularly-labeled structure of the new path-encoded qubits to the structure needed for the post-selected CZ.
We also discuss the optical swap gate that allows the connection of non-nearest neighbor path-encoded qubits. In this way, we can deterministically exchange the locations of the qubits and execute controlled quantum gates between any path-encoded qubits. 
Next, by truncating the auxiliary waveguides after any post-selected CZ, we find that it is possible to cascade this optical gate when it acts on different pairs that share only one qubit.
Finally, we show the Bell state and the Greenberger-Horne-Zeilinger (GHZ) state generation circuits implementing the regular structure, the cascading procedure of post-selected CZ and the optical swap.
\end{abstract*}

\section{Introduction}
\label{sec:intro}
In the last decades, quantum computing (QC)~\cite{nielsen_chuang_2010,bennett2000quantum} has been established as a new computational paradigm, whose logic and rules obey the quantum principles.
Any quantum algorithm is composed of the preparation, manipulation and measurement stages.
The requirements to translate QC from high-level language to its concrete realization are given by DiVincenzo's criteria~\cite{divincenzo_physical_2000}.
Thus, these conditions are used to select suitable platforms to build a scalable and universal quantum computer.

Among different candidates, quantum photonics offers an integrated architecture that can be scaled up and satisfies DiVincenzo's criteria potentially at room temperature and without strict environmental conditions.
In quantum photonics, physical qubits are given by the photons, or more explicitly by their degrees of freedom.
To define photonic qubits, there are different options: time-bin, path, polarization, angular momentum etc~\cite{obrien_optical_2007,tan_resurgence_2019}.
For each choice, we need to operationally establish the fundamental building blocks for the preparation, manipulation and measurement stages, following DiVincenzo's criteria.
In integrated photonics, the spatial degree of freedom is mostly used because of well-established and compact technology that allows the manipulation of single-qubit states.
Single-qubit operators in path encoding can be implemented by a Mach-Zehnder interferometer (MZI), whose building blocks are beam splitters and phase shifters~\cite{cerf_optical_1998,adami_quantum_1999}.
Moreover, enlarging the number of paths for the photons makes possible to arrange $d$-dimensional qubits, or qudits, with the same resources. 
Indeed, the manipulation of a qudit is given by a network of MZIs that executes a generic unitary transformation for a $d$-dimensional vector. 
It is well-known that there are two universal schemes able to make such an operation: Reck~\cite{reck_experimental_1994} and Clements~\cite{clements_optimal_2016} schemes.
To realize a universal photonic two-qubit gate~\cite{divincenzo_two-bit_1995,gottesman_heisenberg_1998},
there are still limitations due to the lack of deterministic structures that operate differently on two-photons states with respect to single-photon states.
Of course, linear transformations do not satisfy this request and, naively, one can think that two-photons gates strictly require some non-linear action.
An example is provided by a Kerr-nonlinearity that operates at the single photon level~\cite{Kerr_milburn}.
However, the breakthrough arrived when it was understood that measurement, post-selection or heralding introduce an effective non-linearity. 
Thus, it is possible to achieve multi-photon gates by looking at a restricted subset of events, selected by specific criteria. 
The price for not using non-linear manipulation is given by a probabilistic gate with a success probability lower than 100\%.
Many successful concepts were demonstrated through linear manipulation of photons~\cite{obrien_demonstration_2003,o2009photonic,laing2010high,li2011reconfigurable}, such as Knill-Laflamme-Milburn protocol~\cite{knill_scheme_2001,Ralph_2001} and post-selection for Controlled-Z gate (CZ)~\cite{postsel_CZ}.
These introductory results show that linear optical quantum computing (LOQC) promises to implement quantum information processing~\cite{o2007optical,rudolph2017optimistic,wang2020integrated,Bartlett2020Universal,Dong_2023}.

In the perspective of large-scale fault-tolerant quantum computing, for example achieved through surface codes~\cite{fowler_surface_2012,google_suppressing_2023}, 
LOQC-based computers have to scale up the capability of computation.
Silicon photonics has demonstrated the potential to integrate large-scale computation protocols~\cite{harris_large-scale_2016, qiang_large-scale_2018}, as well as gate-based quantum computing schemes based on simple optical components~\cite{harris_linear_2018, bernard_photonic_2022, gemma_optical_2022}.
However, the scalability condition requires a regular qubit structure and the possibility of connecting all the qubits.
Two-qubit gates like the post-selected CZ gate do not satisfy these requirements, and therefore represent an obstacle to scalability.
In fact, even though the post-selected CZ gate demonstrates the correct two-qubit operation, the currently implemented structure has trouble in connecting two qubits that are not located in their nearest neighbours.
As we discuss here, these problems can be solved by tailored photonic structures based on reconfigurable photonic networks which enable connecting qubits and arranging a regular structure.
In addition, we explain how to solve the issues related to the connectivity and the cascading of post-selected CZ gates for two different pairs of qubits that share only one qubit.

The paper is organized as follows.
Section~\ref{sec:loqc} reviews the linear optics for quantum computation.
We focus on the path encoding in quantum photonics to define the qubit starting from dual-rail representation. Then, we define single-qubit quantum gates by using linear optical components.
In the same way, we also explain the photonic implementation of CZ gate, which is achieved through post-selection and MZIs as fundamental ingredients.
Based on these backgrounds, Section~\ref{sec:method} describes our proposals for scalability in LOQC on the integrated photonic platform.
Firstly, we describe the regularly-labeled structure of waveguides.
Secondly, we present the optical swap gates, which allow the use of the regular structure or change the location of path-encoded qubits, acting as a generic SWAP quantum gate.
As applied examples, Section~\ref{sec:example} shows the Bell state and Greenberger-Horne-Zeilinger (GHZ) state generation circuits.
Lastly, in Sec.~\ref{sec:conclusion} we summarize the results and present perspectives and problems.

\section{LOQC system}
\label{sec:loqc}
In the following subsections, we start by defining operationally the qubit in a photonic integrated circuit (PIC) by employing the spatial degrees of freedom of the photons. 
After that, we show how to describe the fundamental building blocks for a photonic quantum computation.
In particular, we present path-encoded single-qubit and two-qubit gates and their implementations.

\subsection{Bosonic qubits and Path encoding}
\label{subsec:qubit}
A physical qubit is the fundamental unit of quantum information defined by two distinct discrete states in any system showing quantum mechanical properties, like superposition.

In quantum photonics, a path-encoded qubit is a bosonic qubit, which is defined by the position of one photon between two waveguides.
In other words, this definition utilizes the position of a photon for the two-level system needed for a qubit.
For two waveguides, denoted by $w_0$ and $w_1$ and one photon propagating in one of the two waveguides, we define the configurations for the path-encoded or dual-rail computational basis
\begin{equation}
    \begin{aligned}
    \vert \mathsf{0} \rangle &\equiv \hat{a}_{w_{0}}^\dagger \vert \Omega \rangle = \vert \textbf{1}, \textbf{0} \rangle_{\left(w_{0}, w_1\right)} \,,\\
    \vert \mathsf{1} \rangle &\equiv \hat{a}_{w_1}^\dagger \vert \Omega \rangle = \vert \textbf{0}, \textbf{1} \rangle_{\left(w_{0}, w_1\right)} \,,
    \end{aligned}
    \label{eq:qubit_assigned}
\end{equation}
where $\hat{a}^\dagger$ are creation operators related to the photon in a particular waveguide, $\vert \Omega \rangle$ is the vacuum state, and $w_0$ and $w_1$ indicates the upper and the lower waveguide, respectively. In the previous equation and along the work, we denote with bold numbers $(\textbf{0}, \textbf{1})$ the occupation numbers and with sans serif numbers $(\mathsf{0}, \mathsf{1})$ the computational basis in the considered set of waveguides.
Then, we express an arbitrary quantum state of one bosonic qubit as
\begin{equation}
    \begin{aligned}
        \vert \psi \rangle &= \left( \alpha \, \hat{a}^\dagger_{w_0} + \beta \, \hat{a}^\dagger_{w_1} \right) \vert \Omega \rangle \\
        &\equiv \alpha \, \vert \mathsf{0} \rangle + \beta \, \vert \mathsf{1} \rangle \,,
    \end{aligned}
    \label{eq:generic_1qstate}
\end{equation}
where the probability amplitudes $\alpha$ and $\beta$ of $\vert \psi \rangle$ satisfies the condition $\vert \alpha \vert^2 + \vert \beta \vert^2 = 1$.

Since one qubit consists of one photon and two waveguides, $n$ photonic path-encoded qubits are given by $2n$ waveguides and $n$ identical photons. Each photon is assigned to a pair of waveguides and such collection of two-level systems determines the set of $n$ qubits. This configuration with one photon in every waveguides' pair defines the qubit structure, that must be preserved during the manipulation and at the measurement stage. 
Thus, post-selection is needed in order to select the events that satisfy the qubit structure.
The dual-rail structure is particularly powerful since losses due to real devices do not preserve the qubit structure and can be easily taken into account by post-selection~\cite{bartolucci_fusion-based_2023}.

The generic path-encoded $n$-qubits state reads
\begin{equation}
    \begin{aligned}
        \vert \Psi \rangle &= \bigotimes_{j=0}^{n-1} \vert \psi \rangle_{j} = \prod_{j=0}^{n-1} \left( \alpha_{j} \, \hat{a}^\dagger_{w_0^{j}} + \beta_{j} \, \hat{a}^\dagger_{w_1^{j}} \right) \vert \Omega \rangle \\
        &\equiv \bigotimes_{j=0}^{n-1}  \left( \alpha_{j} \, \vert \mathsf{0} \rangle_j + \beta_{j} \, \vert \mathsf{1} \rangle_j \right) \,,
    \end{aligned}
    \label{eq:generic_nqstate}
\end{equation}
where the index $j$ denotes the $j$-th pair of waveguides which is assigned to the the $j$-th path-encoded qubit and it holds $\vert \alpha_{j} \vert^2 + \vert \beta_{j} \vert^2 = 1$.

Any transformation of single path-encoded qubits can be simply achieved by one MZI (see next subsection for details). The MZIs for single-qubit transformations connect the two waveguides belonging to the same photonic qubit. Consequently, in the ideal case, they preserve the qubit structure since the MZI is a $2\times2$ optical component, i.e. it operates within the single waveguides' pair. In the case of real devices, losses decrease the number of photons and therefore create events outside the qubit structure that can be distinguished.

In addition to single-qubit operations, we need only one controlled two-qubit operation to achieve a set of universal gates.
In order to introduce two-qubit gates, the pairs of waveguides need to interact and preserve the qubit structure.
To achieve this, we can consider suitable networks of MZIs: in this way we can exploit the same reconfigurable resource used for single-qubit operations.
Given a generic set of $m$ waveguides, the universal transformation can be implemented by a $m\times m$ network of MZIs~\cite{reck_experimental_1994,clements_optimal_2016}. The term "universal" does not mean that such a network can execute QC universal gates like CZ and CNOT. It means that any input single-photon state propagating in the $m$-modes circuit can be mapped to any output single-photon state. More rigorously, the MZI network implements a generic unitary transformation for $m$-dimensional complex vector, which represents the probability amplitudes of the single photon propagating in $m$ paths. For example, given a single photon inserted in the $j$-th input, the transformation is
\vspace{-0.2cm}
\begin{equation}
    \hat{a}^\dagger_{j}
    \to
    \sum_{k=1}^m \,u_{jk} \,\hat{a}^\dagger_{k}
    \implies
    \vert \overbrace{\textbf{0}\ldots\textbf{0}}^{j-1} \textbf{1}   ,\textbf{0}\ldots\textbf{0} \rangle_{\left(w_1\ldots w_m\right)}
    \to 
    \sum_{k=1}^m \,u_{jk} \vert \overbrace{\textbf{0}\ldots\textbf{0}}^{k-1} \textbf{1} \,\textbf{0}\ldots\textbf{0}  \rangle_{\left(w_1\ldots w_m\right)}
\end{equation}
where $\hat{a}^\dagger$s are the creation operator of the network modes $\left(w_1\ldots w_m\right)$ and $u_{jk}$ are the matrix elements that describe the action of the MZIs network. The previous equation tells us that the input state with a photon in the $j$-th waveguide is transformed into a generic superposition state of the photon in the set of $m$ waveguides.
When we insert more than one photon in a generic network, even if the transformation is linear, the interference between indistinguishable outputs produces not trivial results. Examples of non-triviality are given by the Hong-Ou-Mandel effect~\cite{hong_measurement_1987} and Boson Sampler~\cite{boson_sampling,boson_sampling_review}. In the first case, two single photons are inserted in the inputs of a balanced beam splitter producing an entangled path-encoded state~\cite{bouchard2020two,branczyk2017hong}, and in the second case, the result is linked to the permanent~\cite{permanent} of the generic unitary transformation represented by the MZI network.

The complexity of MZI networks suggests that multi-photon gates can be realized by suitable choices of the MZIs' phases and by post-selection operations to preserve the desired qubit structure.
Naively, one can think that a $4\times4$ MZI network connecting two pairs of waveguides with two single photons could implement a universal two-qubit gate such as the CNOT and CZ gates.
Unfortunately, as we demonstrate in Appendix~\ref{app:4x4imp}, this does not occur. Larger MZI matrices are needed to implement a controlled two-qubit gate. Noteworthy, from the point of view of resources and within the gate-based approach, the post-selected CZ gate~\cite{postsel_CZ} is the most efficient way to implement the CZ operation. 

\begin{figure}[t]
    \centering
    \includegraphics[width=0.70\textwidth]{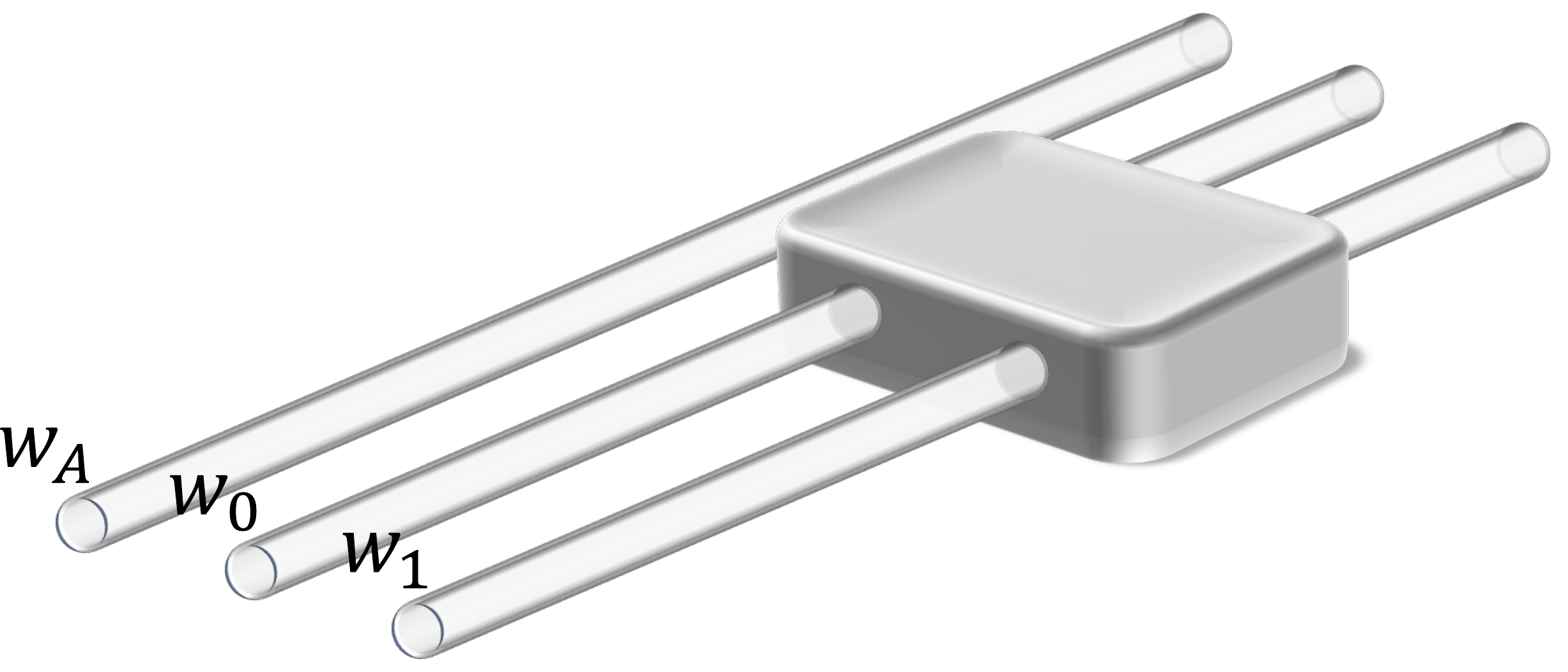}
    \caption{Graphical representation of three waveguides that define the chosen photonic path-encoded qubit. The upper waveguide $w_A$ is the auxiliary waveguide needed for CZ gates, and the doublet $( w_{0}, w_{1} )$ composes the computational basis states of the path-encoded photonic qubit. 
    We also report a box that connects the waveguides $( w_{0}, w_{1} )$ and represents a generic single-qubit gate implemented by a MZI.}
    \label{fig:qubit_3waveguides}
\end{figure}

Therefore, we use this gate as a benchmark. The realization of the post-selected CZ gate involves two additional waveguides which are denoted as auxiliary waveguides. Thus, six total waveguides and two photons are required to implement such a CZ gate between two qubits. 
At the end of the operation, it is also necessary to check that the qubit structure is preserved. This post-selection operation makes probabilistic the correct execution of the gate.
The configuration with six waveguides for the CZ gate between two qubits suggests that an auxiliary waveguide can be assigned to each pair of waveguides that defines one dual-rail qubit. In Fig.~\ref{fig:qubit_3waveguides} we report a graphical representation of the chosen photonic qubit.
Therefore, the path-encoded qubit is re-defined by one photon that propagates in three waveguides, denoted as $(w_{A}, w_{0}, w_{1})$. 
When we consider $n$ qubits, we denote their triplet of waveguides as $( w_{A}^{j}, w_{0}^{j}, w_{1}^{j} )$. 
The qubit structure requires that only one photon is present in each doublet $( w_{0}^{j}, w_{1}^{j} )$, which represents the path-encoded computational basis, and no photon in $w_A^{j}$ at the input and output of every gate.
This is trivially satisfied by single-qubit gates that operate within every doublet $( w_{0}^{j}, w_{1}^{j} )$, but it is not guaranteed by the post-selected CZ. As it is explained later more explicitly, even in the ideal case with no losses, the post-selected CZ gives rise to events with photons in the auxiliary waveguides and zero or two photons in the waveguides' pair corresponding to the same qubit. In the real case with losses, we have also to discard events with a decreased number of photons with respect to the initial ones.

To summarize, whenever the output is composed of at least one photon in any auxiliary waveguide, by more than one photon in any doublet $( w_{0}^{j}, w_{1}^{j} )$ or by less than $n$ photons due to losses, we discard the event.
We note that it is possible to make the events that involve losses and the presence of photons in the auxiliary waveguides equivalent by truncating, or equivalently not connecting, the auxiliary waveguides at the output of every post-selected CZ gates. In this way, the post-selection at the output is simplified since no photons are coming from such waveguides by construction. We denote this procedure of interrupting or not connecting the auxiliary waveguides after any post-selected CZ gate as the {\it truncation trick}.

Once the path-encoded qubit structure is set, we can easily define how to initialize and measure the qubits.
The initialization stage is straightforward, since we can set the initial state simply by injecting single-photon states in the relative waveguides. For example, if we want to start with the state $\bigotimes_{j=0}^{n-1} \vert \mathsf{0} \rangle$, we prepare the initial configuration with one photon is each waveguide $w_{0}^{j}$.
The measurement stage is implemented by single-photon detectors and the correct events are given by the ones that satisfy the qubit structure. Thus, if we have $n$ qubits, the events that must be considered are only those where each of the $n$ photons is measured between every pair of waveguides $( w_{0}^{j}, w_{1}^{j} )$ with the corresponding assignment given in Eq.~\eqref{eq:qubit_assigned}. 

Concerning the manipulation stage, we exploit the tunability and versatility of MZI networks by embedding the path-encoded qubits gates in the Reck~\cite{reck_experimental_1994} and Clements~\cite{clements_optimal_2016} schemes. In particular, since our choice maps $n$ qubits in $3n$ waveguides, we consider a MZI network with $3n$ modes to implement the desired operations on the qubits.

\subsection{Single-qubit operators}
\label{subsec:singleop}
This subsection describes single-qubit quantum gates in LOQC. 
To achieve this, we introduce the fundamental building blocks, needed to linearly manipulate photon states inside a PIC.

The first building block is the 50:50 beam splitter (BS)~\cite{wang2020integrated}.
This component has two inputs and two outputs and can be realized by utilizing two optical phenomena: evanescent wave or self-imaging.
In the first case, we have directional couplers, while in the second case multimode-interferometer-based devices (MMIs).
The action of BS is well described by its name: if light is injected in one of the two entries, the outputs' intensities are half of the initial one.
BS can be characterized by an unbalanced condition between the two outputs and insertion loss, which is the amount of power that is lost in the photonic device.
We can describe this situation as
\begin{equation}
    {\rm BS}_{\rm gen}(t,r) =
    \begin{pmatrix}
        t & {\rm i}\,r \\
        {\rm i}\,r & t
    \end{pmatrix} \,,
    \label{eq:beam_general}
\end{equation}
where the label "gen" stands for generic and $t$ and $r$ are the amplitudes of the two outputs and they are denoted as transmittance and reflectance. Unbalance means that $|t| \ne |r|$ and insertion loss implies that  $|t|^2 + |r|^2 < 1$.
If the BS is ideal, the corresponding matrix is 
\begin{equation}
    {\rm BS} = \frac{1}{\sqrt{2}}
    \begin{pmatrix}
        1 & \pm{\rm i} \\
        \pm{\rm i} & 1
    \end{pmatrix} \,,
    \label{eq:beam_ideal}
\end{equation}
modulo a global phase. The plus sign corresponds to the MMIs case, while the minus sign to the directional coupler case.
It is interesting to note that the conditions to be balanced, lossless and unitary fixes all the relative phases of the device.

The second component is given by the phase shifter (PS)~\cite{wang2020integrated}.
This component is used to change the relative phase of different spatial degrees of freedom belonging to the same dual-rail qubit.
When photons propagate in a waveguide for a distance $L$, they accumulate an overall phase given by $2\pi\,L\,n_{\rm eff}/\lambda$.
The effective index $n_{\rm eff}$ is a parameter that depends on all the characteristics of the waveguide: in particular if all the waveguides are identical, no relative phase can be induced.
However, by changing external parameters like temperature, pressure, etc., we can locally vary the effective index.
Therefore, if we have a couple of waveguides and one photon, PSs on each path introduce two different phases.
This configuration involves two inputs and two outputs and it can be described by the following matrix
\begin{equation}
    {\rm PS}(\boldsymbol{\theta}) = 
    \begin{pmatrix}
        {\rm e}^{{\rm i}\theta_1} & 0 \\
        0 & {\rm e}^{{\rm i}\theta_2}
    \end{pmatrix} \,,
    \label{eq:phase_shift}
\end{equation}
where $\boldsymbol{\theta} = (\theta_1,\theta_2)$.
Note that after this operation, we can measure only the relative phase $\theta_1-\theta_2$, if we look just at one waveguides' pair.

These two optical linear components constitute the building blocks to manipulate photon states through a MZI.
\begin{figure}[t]
    \centering
    \includegraphics[width=0.70\textwidth]{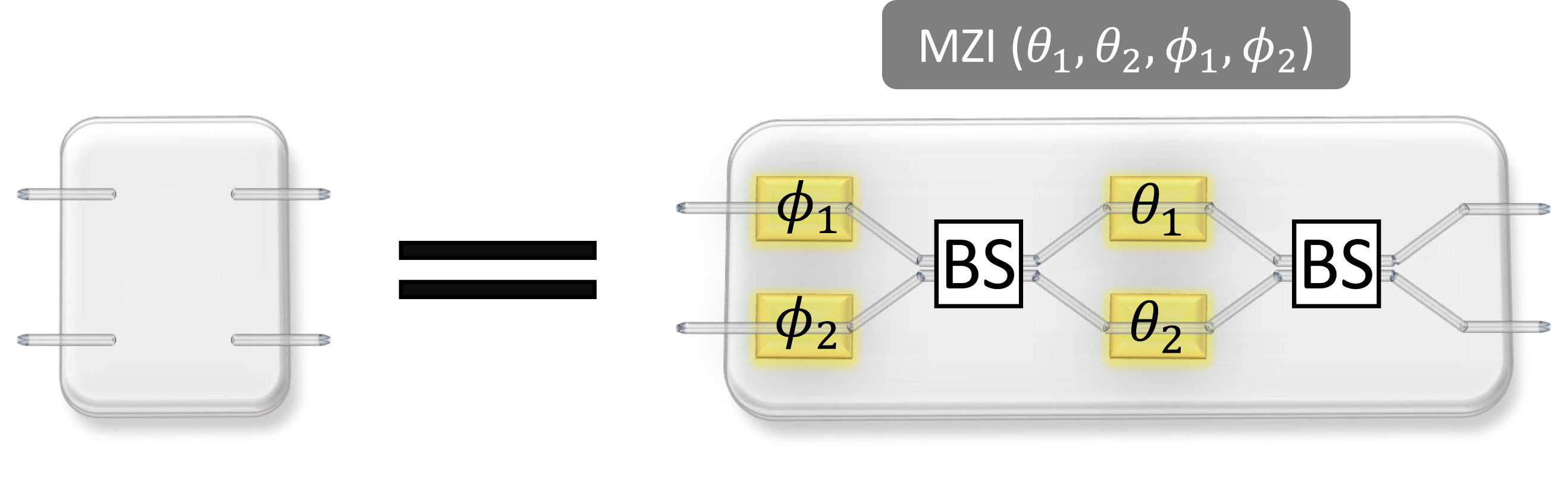}
    \caption{One Mach-Zehnder interferometer unit. 
    This building block is composed of two beam splitters (white square boxes labeled with BS) and four phase shifters (yellow square boxes labeled with $\theta$s and $\phi$s), located at the inputs and in the internal arms.
    In all the following images, we represent the MZI photonic unit with the light grey box.}
    \label{fig:mziunit}
\end{figure}
As it is shown in Fig.~\ref{fig:mziunit}, two BSs and two pairs of PSs constitute a MZI.
Consequently, the unitary matrix for a MZI is 
\begin{equation}
    \begin{split}
    U_{\rm MZI}( \boldsymbol{\theta} )  &\equiv {\rm BS} \cdot {\rm PS}(\boldsymbol{\theta}) \cdot {\rm BS} \cdot {\rm PS}(\boldsymbol{\phi}) \\
    &= {\rm i}\, {\rm e}^{{\rm i}(\theta_1+\theta_2+2\phi_2)/2} 
    ~~~~
    \begin{pmatrix}
        {\rm e}^{{\rm i}(\phi_1-\phi_2)}\sin((\theta_1-\theta_2)/2) & \cos((\theta_1-\theta_2)/2) \\
        {\rm e}^{{\rm i}(\phi_1-\phi_2)}\cos((\theta_1-\theta_2)/2) & -\sin((\theta_1-\theta_2)/2)
    \end{pmatrix} \,,
    \end{split}
\label{eq:MZI_matrix}
\end{equation}
where $\boldsymbol{\theta} = (\theta_1,\theta_2)$ and $\boldsymbol{\phi} = (\phi_1,\phi_2)$.
Note that Eq.~\eqref{eq:MZI_matrix} has the same representation of Eq.~\eqref{eq:beam_general}, that describes a generic beam splitter. This means that the MZI is equivalent to a beam splitter with tunable parameters.

Moreover, Eq.~\eqref{eq:MZI_matrix} implies that we can execute single-qubit gates. 
We list some commonly used single-qubit gates in Table.~\ref{tab:quantumgates}.
Most of the quantum gates can be evaluated by the phases' configuration depicted in Fig.~\ref{fig:mziunit}. 
The rotation around the $z$-axis of the Bloch sphere $R_z$ is achieved by setting $(\theta_1-\theta_2,\phi_1-\phi_2)=(\pi,\pi-\delta)$.
However, the other rotations $R_x$ and $R_y$ require an additional pair of PSs at the output of the MZI. In this way, we have an extended version of the MZI, whose matrix reads 
\begin{equation}
    U_{\rm ext\,\,MZI} \equiv 
    {\rm PS}(\phi_3,\phi_4) \cdot U_{\rm MZI}(\boldsymbol{\theta}, \boldsymbol{\phi}) \,.
\end{equation}
Through this addition, $R_x$ and $R_y$ can be achieved by the phases configurations $(\theta_1-\theta_2,\phi_1-\phi_2,\phi_3-\phi_4)$:
$(\pi-\delta,\pi/2,\pi/2)$ and $(\pi-\delta,0,\pi)$, respectively.
The previous matrix representation can be used following the recipe given in App.~\ref{app:mxm}. Therefore, to keep a compact notation, we create a vector with all the creation operators associated with the inputs and simply multiply such vector by the matrix that describes the desired device.
In the case of a MZI, the transformation can be written as
\begin{equation}
\begin{pmatrix}
        \hat{a}_{w_{0}}^\dagger \\
        \hat{a}_{w_{1}}^\dagger 
    \end{pmatrix}
    \to \,\, U_{\rm MZI}(\boldsymbol{\theta}, \boldsymbol{\phi})^{-1}
    \cdot
    \begin{pmatrix}
        \hat{a}_{w_{0}}^\dagger \\
        \hat{a}_{w_{1}}^\dagger
    \end{pmatrix}\,,
    \label{eq:MZItra}
\end{equation}
and this can be generalized to a generic $m$ mode network by creating a vector containing all the creation operators.

The MZI structure can be easily embedded into the three-waveguide structure as illustrated in Fig.~\ref{fig:qubit_3waveguides}. The auxiliary waveguide is left alone and the waveguides corresponding to the computational basis are connected through the MZI.
One can follow the procedure reported in App.~\ref{app:mxm} to find the matrix corresponding to the single-qubit manipulation implemented by a MZI for the three-waveguide structure.

\begin{table}[t]
\centering
\caption{Collection of commonly used single qubit gates evaluated by specific phase values. Through changing two parameters $\theta$ and $\phi$ in Eq.~\eqref{eq:MZI_matrix}, quantum gates are determined, modulo a global phase.}\label{tab:quantumgates}
\resizebox{0.65\textwidth}{!}{
\begin{tabular}{l l l}
\toprule
\addlinespace[0.3em]
\multicolumn{1}{c}{Quantum gates} & \multicolumn{1}{l}{Matrix form} & MZI phases  $(\theta_1-\theta_2,\phi_1-\phi_2)$ 
\\
\addlinespace[0.3em] \midrule \addlinespace[0.3em]
Identity        & \footnotesize$\begin{pmatrix} 1 & 0 \\ 0 & 1 \end{pmatrix}$                    &  $(\pi,\pi)$
\\
\addlinespace[0.3em] \hline \addlinespace[0.3em]
Pauli-X         & \footnotesize$\begin{pmatrix} 0 & 1 \\ 1 & 0 \end{pmatrix}$                    &      $(0,0)$
\\
\addlinespace[0.2em] \hline \addlinespace[0.2em]
Pauli-Y         & \footnotesize$\begin{pmatrix} 0 & -{\rm i}\\ {\rm i}& 0 \end{pmatrix}$                   &           $(0,\pi)$
\\
\addlinespace[0.2em] \hline \addlinespace[0.2em]
Pauli-Z         & \footnotesize$\begin{pmatrix} 1 & 0 \\ 0 & -1 \end{pmatrix}$                   &          $(\pi,0)$
\\
\addlinespace[0.2em] \hline \addlinespace[0.2em]
Hadamard        & \footnotesize$\frac{1}{\sqrt{2}}\begin{pmatrix} 1 & 1 \\ 1 & -1 \end{pmatrix}$ &        $(\pi/2,0)$
\\
\addlinespace[0.2em] \hline \addlinespace[0.2em]
T               & \footnotesize$\begin{pmatrix} 1 & 0 \\ 0 & {\rm e}^{{\rm i}\pi/4} \end{pmatrix}$     &         $(\pi,-\pi/4)$
\\
\bottomrule
\end{tabular}
}
\end{table}
\subsection{Two-qubit operators}
\label{subsec:twoop}
\begin{figure}[t]
    \centering
    \includegraphics[width=0.60\textwidth]{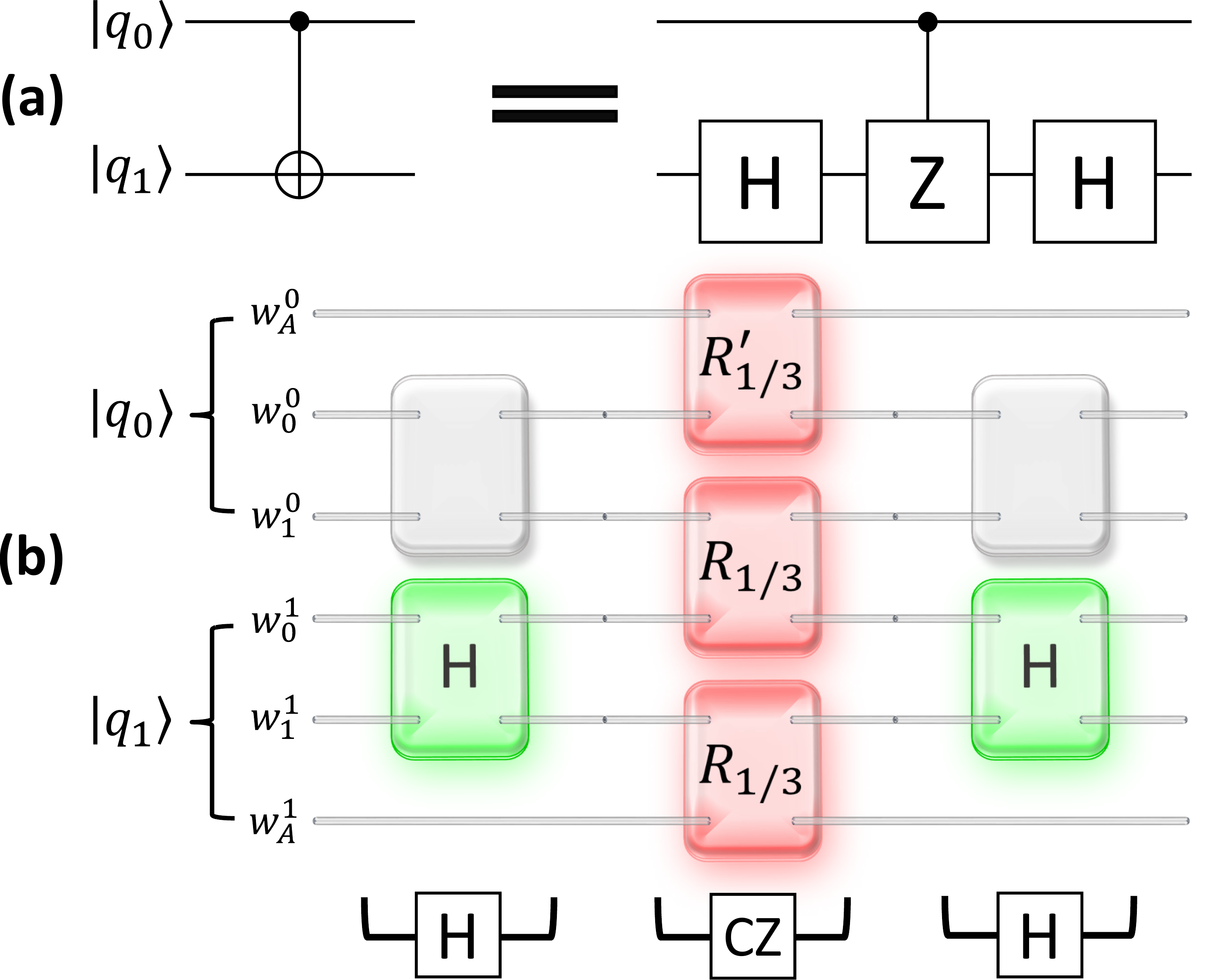}
    \caption{
    CNOT implementation through MZI units for path-encoded photonic qubits.
    (a) Gate decomposition of CNOT with the combination of H and CZ.
    (b) CNOT implementation with MZIs. 
    $(w_{A}^{0}, w_{A}^{1})$ are the ancillary waveguides, $(w_{0}^{0}, w_{1}^{0})$ and $(w_{0}^{1}, w_{1}^{1})$ indicate the waveguides of the control and target qubits, respectively.
    Post-selected CZ is located in the center of the MZI network, and Hadamard gates needed for CNOT are positioned on the left and right of the post-selected CZ.
    The MZIs are set in such a way to have the transformation H, $R_{1/3}$ and $R'_{1/3}$: the values for the phases can be found in Table.~\ref{tab:quantumgates}, and in Eqs.~\eqref{R1/3}. The MZIs represented with light grey boxes are set in the identity configuration.}
    \label{fig:cnot}
\end{figure}

Among the various CZ implementations in LOQC, we embrace the post-selected method, implemented in previous studies~\cite{obrien_demonstration_2003, mower_high-fidelity_2015, lee_controlled-not_2022}.
The main idea of the post-selected CZ consists of adding two auxiliary waveguides to the four waveguides belonging to the two dual-rail qubits.
Then, the post-selected CZ gate is implemented by a layer of three beam splitters with $1/3$ transmittance, which we denote as $1/3$-BS.
Additionally, the CNOT gate can be realized by inserting the $1/3$-BS between two 50:50 BSs acting on the waveguides relative to the target qubit. 50:50 BS performs the analogous role as the H gate by the equality of gate decomposition~\cite{barenco_elementary_1995, zhou_methodology_2000}. 
Fig.~\ref{fig:cnot} graphically reports the structure of the post-selected CNOT, composed of post-selected CZ and single-qubit gates with MZIs.
Here, we detail the post-selected CZ and CNOT.

First of all, we represent two-qubit states $\vert q_0 \rangle$ and $\vert q_1\rangle$ in the computational basis $\{ | \mathsf{0} \rangle\,, | \mathsf{1} \rangle \}$ through the path-encoded correspondence given in Eq.~\eqref{eq:generic_nqstate}. 
The basis states read as
\begin{equation}
\begin{split}
    \vert \mathsf{00} \rangle \equiv \hat{a}_{w_{0}^{0}}^\dagger\hat{a}_{w_{0}^{1}}^\dagger \vert \Omega \rangle = \vert \textbf{1},\textbf{0},\textbf{1},\textbf{0} \rangle_{\left(w_{0}^{0}, w_{1}^{0}, w_{0}^{1}, w_{1}^{1}\right)} \,, \\
    \vert \mathsf{01} \rangle \equiv \hat{a}_{w_{0}^{0}}^\dagger\hat{a}_{w_{1}^{1}}^\dagger \vert \Omega \rangle = \vert \textbf{1},\textbf{0},\textbf{0},\textbf{1} \rangle_{\left(w_{0}^{0}, w_{1}^{0}, w_{0}^{1}, w_{1}^{1}\right)} \,, \\
    \vert \mathsf{10} \rangle \equiv \hat{a}_{w_{1}^{0}}^\dagger\hat{a}_{w_{0}^{1}}^\dagger \vert \Omega \rangle = \vert \textbf{0},\textbf{1},\textbf{1},\textbf{0} \rangle_{\left(w_{0}^{0}, w_{1}^{0}, w_{0}^{1}, w_{1}^{1}\right)} \,, \\
    \vert \mathsf{11} \rangle \equiv \hat{a}_{w_{1}^{0}}^\dagger\hat{a}_{w_{1}^{1}}^\dagger \vert \Omega \rangle = \vert \textbf{0},\textbf{1},\textbf{0},\textbf{1} \rangle_{\left(w_{0}^{0}, w_{1}^{0}, w_{0}^{1}, w_{1}^{1}\right)} \,,
\end{split}
\label{eq:basis}
\end{equation}
where the two pairs of waveguides associated with the qubits are $(w_{0}^{0}, w_{1}^{0})$ and $(w_{0}^{1}, w_{1}^{1})$. These two doublets are put between two auxiliary waveguides $\left(w_{A}^{0}, w_{A}^{1}\right)$, as shown in the input and output of scheme~\ref{fig:cnot}(b).
Note that this structure is not regular.

Then, we introduce the operators $R_{1/3}$ and $R'_{1/3}$ that describe the $1/3$-BSs present in the post-selected CZ.
The values of $\theta$s and $\phi$s and the corresponding $R_{1/3}$ and $R'_{1/3}$ reads as
\begin{equation}
    \begin{aligned}
    &\begin{rcases}
        & \phi_1 - \phi_2 =\pi  \\
        & \theta_1-\theta_2 = 2 \arcsin \frac{1}{\sqrt{3}}
    \end{rcases}  \Rightarrow 
    R_{1/3} =  \frac{1}{\sqrt{3}} 
    \begin{pmatrix}
        -1 & \sqrt{2} \\
        -\sqrt{2} & -1
    \end{pmatrix} \,,\\
        &\begin{rcases}
            & \phi_1 - \phi_2 = 0  \\
            & \theta_1-\theta_2 = -2 \arcsin \frac{1}{\sqrt{3}}\\
        \end{rcases} \Rightarrow
        R'_{1/3} =  \frac{1}{\sqrt{3}}
        \begin{pmatrix}
            -1 & \sqrt{2}\\
            \sqrt{2} & 1
        \end{pmatrix} \,,
    \end{aligned}
     \label{R1/3}
\end{equation}
modulo a global phase.
The state evolution associated with the post-selected CZ can be written as
\begin{equation}
    \overline{\rm V}_2
    \to
    \overline{\rm CZ}_{\rm ps}
    \cdot
    \overline{\rm V}_2 \,,
    \label{eq:recipeCZ}
\end{equation}
where
\begin{equation}
\begin{split}
    \overline{\rm V}_2^{\rm T} &\equiv 
    \left(
        \hat{a}_{w_{A}^{0}}^\dagger,
        \hat{a}_{w_{0}^{0}}^\dagger,
        \hat{a}_{w_{1}^{0}}^\dagger,
        \hat{a}_{w_{0}^{1}}^\dagger,
        \hat{a}_{w_{1}^{1}}^\dagger,
        \hat{a}_{w_{A}^{1}}^\dagger
    \right) \,,\\
\overline{\rm CZ}_{\rm ps} &\equiv 
\frac{1}{\sqrt{3}}
\footnotesize
\begin{pmatrix}
    -1 & \sqrt{2} &  0 & 0 & 0 & 0 \\
    \sqrt{2} & 1 & 0  & 0 & 0 & 0 \\
    0 & 0 & -1 & -\sqrt{2} & 0 & 0 \\
    0 & 0 & \sqrt{2} & -1 & 0 & 0 \\
    0 & 0 & 0 & 0 & -1 & -\sqrt{2} \\
    0 & 0 & 0 & 0 & \sqrt{2} & -1 \\
\end{pmatrix} \,,
\end{split}
\end{equation}
the superscript "T" denotes the transpose and the bar is inserted as a reminder of the non-regularly-labeled structure of the gate. Note also that $\overline{\rm CZ}_{\rm ps}$ is the inverse matrix of the block matrix composed by $R_{1/3}$ and $R'_{1/3}$, following the procedure explained around Eq.~\eqref{eq:MZItra} and in App.~\ref{app:mxm}.

The elements corresponding to the auxiliary waveguides in $\overline{\rm V}_2$ are prepared to have zero probability, since there is no photon in the two ancillary waveguides $\left(w_{A}^{0}, w_{A}^{1}\right)$.
Indeed, the generic initial state is
\begin{equation}
\begin{split}
    \vert \Psi \rangle_{\rm in} &= \vert q_0 \rangle_{\rm in} \otimes \vert q_1 \rangle_{\rm in} \,, \\
    \vert q_0 \rangle_{\rm in} &=
    \left( \alpha_{0} \,\hat{a}_{w_{0}^{0}}^\dagger + \beta_{0} \,\hat{a}_{w_{1}^{0}}^\dagger \right)\vert \Omega \rangle 
    \equiv
     \alpha_{0} \, \vert \mathsf{0} \rangle_0 + \beta_{0} \, \vert \mathsf{1} \rangle_0 \,, \\
    \vert q_1 \rangle_{\rm in} &= \left( \alpha_{1} \,\hat{a}_{w_{0}^{1}}^\dagger + \beta_{1} \,\hat{a}_{w_{1}^{1}}^\dagger \right) \vert \Omega \rangle
     \equiv
     \alpha_{1} \, \vert \mathsf{0} \rangle_1 + \beta_{1} \, \vert \mathsf{1} \rangle_1 \,,
    \end{split}
\label{eq:generic_2qstate}
\end{equation}
where we follow the prescriptions given in Sec.~\ref{subsec:qubit} and the probability of finding a photon in the auxiliary waveguides is zero.

We assign the control to the first qubit and the target to the second one, called $\vert q_0 \rangle$ and $\vert q_1 \rangle$ in Fig.~\ref{fig:cnot}.
Applying the linear transformation in Eq.~\eqref{eq:recipeCZ}, the state $\vert \Psi \rangle_{\rm in}$ evolves to $\vert \Psi \rangle_{\rm out}$, which reads modulo global phase as
\begin{equation}
    \begin{aligned}
        \vert \Psi \rangle_{\rm out}
        =& -\frac{1}{3} \left(   \alpha_{0} \alpha_{1} \,\hat{a}_{w_{0}^{0}}^\dagger \hat{a}_{w_{0}^{1}}^\dagger + \alpha_{0}\beta_{1}\,\hat{a}_{w_{0}^{0}}^\dagger\hat{a}_{w_{1}^{1}}^\dagger 
        + \beta_{0} \alpha_{1}\,\hat{a}_{w_{1}^{0}}^\dagger \hat{a}_{w_{0}^{1}}^\dagger - \beta_{0} \beta_{1} \,\hat{a}_{w_{1}^{0}}^\dagger \hat{a}_{w_{1}^{1}}^\dagger\right) \vert \Omega \rangle 
        +\ldots
    \end{aligned}
    \label{eq:cz_hp}
\end{equation}
where the dots contain all the terms that do not preserve the initial qubit structure. In particular, among these terms we find events with 
\begin{enumerate}
    \item one photon in $w_{A}^{0}$ and one photon in the doublet $\left( w_{0}^{0}, w_{1}^{0}\right)$ or $\left( w_{0}^{1}, w_{1}^{1}\right)$,
    \item one photon in $w_{A}^{1}$ and one photon in the doublet $\left( w_{0}^{0}, w_{1}^{0}\right)$ or $\left( w_{0}^{1}, w_{1}^{1}\right)$,
    \item two photons in the auxiliary waveguides $\left( w_{A}^{0}, w_{A}^{1}\right)$,
    \item two photons in the doublet $\left( w_{0}^{0}, w_{1}^{0}\right)$ or $\left( w_{0}^{1}, w_{1}^{1}\right)$.
\end{enumerate}
The post-selection discards these events. In the presence of losses, we also discard events with one or no photon in the six waveguides set.

Finally, we can rewrite the Eq.~\eqref{eq:cz_hp} using the assignment with computational basis, given Eq.~\eqref{eq:basis}:
\begin{equation}
    \begin{aligned}
        \vert \Psi \rangle_{\rm out} \propto \alpha_{0} \alpha_{1} \vert \mathsf{00} \rangle + \alpha_{0} \beta_{1} \vert \mathsf{01} \rangle + \beta_{0} \alpha_{1} \vert \mathsf{10} \rangle - \beta_{0} \beta_{1} \vert \mathsf{11} \rangle  \,,
    \end{aligned}
    \label{eq:cz_corr}
\end{equation}
where we neglect the contributions that are non-preserving the qubit structure.

Apart from a global phase, the result shows the correct CZ operation on $\vert \Psi \rangle_{\rm in}$: there is a sign change when both qubits are in the state $\vert \mathsf{1} \rangle$. Once we have the CZ gate, the CNOT gate is simply achieved by performing the H gate on the target qubit before and after the CZ. The H gate is simply achieved by setting phase parameters denoted in Table~\ref{tab:quantumgates} to a MZI under the 50:50 BS configuration. The CNOT gate is shown in Fig.~\ref{fig:cnot} and its action can be described as 
\begin{equation}
    \overline{\rm CNOT}_{\rm ps} = \left( \mathbf{1}\otimes {\rm H} \right) \cdot \overline{\rm CZ}_{\rm ps} \cdot \left( \mathbf{1}\otimes {\rm H} \right) \,,
\end{equation}
where the transformation $\mathbf{1}\otimes {\rm H}$ represents the action of H gate on the second qubit. 

If we denote by $\hat{\rm P}$ the operator that projects on the qubit structure and equivalently eliminates the terms contained in the dots of Eq.~\eqref{eq:cz_hp}, the success probability $\mathbb{P}_{\rm succ}$ of $\hat{\rm P} \vert \Psi \rangle_{\rm out}$ is
\begin{equation}
    \mathbb{P}_{\rm succ} = {}_{\rm out}\langle \Psi \vert \hat{\rm P} \vert \Psi \rangle_{\rm out} = \frac{1}{9}\,,
    \label{eq:cz_succ}
\end{equation}
whose value is exactly the square of the coefficient in front of the part of $\vert \Psi \rangle_{\rm out}$ that preserves the qubit structure. In the case of losses, this value decreases.

It is well known that this implementation of the CZ cannot be cascaded: this means that if this operation is applied twice to the same pair of qubits or to different pairs of qubits that share only one qubit, the correct result cannot be achieved by the post-selection procedure. 
As explained in Sec.~\ref{subsec:qubit}, the {\it truncation trick} makes the events with at least one photon in the auxiliary waveguides equivalent to the event where we lose at least one photon because of losses. This is achieved by truncating the auxiliary waveguides at the output of the post-selected CZ. The {\it truncation trick} can be described by the following non-unitary matrix for two path-encoded qubits
\begin{equation}
    \hat{\rm P}_{\rm aux} = 
    \begin{pmatrix}
    0 & 0 & 0 & 0 & 0 & 0  \\
    0 & 1 & 0 & 0 & 0 & 0  \\
    0 & 0 & 1 & 0 & 0 & 0  \\
    0 & 0 & 0 & 0 & 0 & 0  \\
    0 & 0 & 0 & 0 & 1 & 0  \\
    0 & 0 & 0 & 0 & 0 & 1 
    \end{pmatrix}\,.
\end{equation}
This form can be easily generalized to any number of qubits.
The action of the previous matrix consists of dropping all the terms with at least one photon in the auxiliary waveguides by default. This operation does not include the post-selection procedure completely, since we still have the events with two photons in the pair of waveguides that corresponds to one of the two path-encoded qubits.
These events can be distinguished by single-photon detectors, but they are detrimental when two post-selected CZ are cascaded on the same pair of qubits, even if we use the {\it truncation trick}.
However, as shown in App.~\ref{app:GHZ}, the {\it truncation trick} makes possible the correct operation of two post-selected CZ gates applied to two pairs of qubits that share only one qubit.

We conclude this subsection by considering Fig.~\ref{fig:unischeme}(a), where we report the Clements~\cite{clements_optimal_2016} and Reck~\cite{reck_experimental_1994} schemes. It is possible to note that the post-selected CZ can be embedded in five different ways in the Clements scheme and only one way in the Reck scheme. Indeed, the essential condition consists of a layer composed of three MZI.
Therefore, a $6\times6$ universal scheme can be set to execute the QC universal set of gates for two path-encoded qubits.

\subsection{Issues in large-scale photonic QC}
The post-selected CZ gate, shown in Fig.~\ref{fig:cnot}(b), has the limitation that does not satisfy scalability, since it involves a non-regularly-labeled structure. Indeed, its structure is composed of the two pairs of waveguides $(w_{0}^{0}, w_{1}^{0})$ and $(w_{0}^{1}, w_{1}^{1})$ corresponding to the two qubits $\vert q_0\rangle$ and $\vert q_1\rangle$, that are located between the auxiliary waveguides $( w_A^0, w_A^1)$.
The waveguides' assignment is not regular and it only permits an operation between nearest-neighbor qubits. This hinders the flexibility of two-qubit operations.
For instance, if we want to sequentially apply CNOT gates on the pairs of qubits $(\, \vert q_0\rangle, \vert q_1\rangle \,)$, $(\, \vert q_1\rangle, \vert q_2\rangle \,)$, the current scheme cannot be applied due to the lack of correspondence between the qubit structure, i.e the path assignment does not match among different sequential post-selected CZ gates.
Therefore, the irregular configuration of waveguides in the post-selected CZ prevents satisfying the scalability property in MZI networks.

Large-scale QC requires quantum operations between non-adjacent qubits.
Suppose that we would operate two-qubit gates on three qubits, $\vert q_0\rangle$, $\vert q_1\rangle$, and $\vert q_2\rangle$, where $(\, \vert q_0\rangle, \vert q_1\rangle \,)$ and $(\, \vert q_1\rangle, \vert q_2\rangle \,)$ are adjacent.
In this situation, we need to exchange the location of $\vert q_0\rangle$ or $\vert q_2\rangle$ with $\vert q_1\rangle$ to connect $\vert q_0\rangle$ and $\vert q_2\rangle$.
The SWAP gate performs this operation and it can be realized by three successive CNOT gates on the same pair of qubits.
However, the CZ implementation, discussed in Sec.~\ref{subsec:twoop}, does not allow to cascade more CZ gates on two qubits, even if we use the {\it truncation trick}.
In integrated photonics, the swapping of spatial degrees of freedom is typically done by waveguide crossings~\cite{bozkurt_linear_2021}. 
This optical component achieves the same operation that a MZI performs in the cross configuration, but in a passive way and with a more compact footprint. SWAP between qubits or waveguides given by the cross configuration, or equivalently Pauli-$X$ gate, is achieved without setting any relative phases in the MZI unit (see Table.~\ref{tab:quantumgates}). This means that also MZIs do not need external power to work in the cross configuration. 
Even though, in a real MZI, spurious phases, coming from imperfections, must be compensated and external power to set the correct configuration is needed, the advantage of using MZI networks to execute the swapping of path degrees of freedom comes from their reconfigurability. 

In the following section, we present the regularly-labeled structure and the optical swap gate to solve the previously reported scalability issues.

\section{Method for scalability}
\label{sec:method}
In a photonic network consisting of MZIs, we describe two criteria to improve the scalability of LOQC: on the implementation level and on the quantum gate level.
We demonstrate the post-selected CZ gate on a regularly-labeled structure with one ancillary waveguide and one dual-rail as data qubit, as shown in Fig.~\ref{fig:qubit_3waveguides}.
We refer to recent work on bulk optics~\cite{meng_deterministic_2022} to introduce an optical SWAP gate, which maps the regular structure to the standard structure needed to execute the post-selected CZ gate.

\subsection{Primitives on implementation level}
\subsubsection{Regular labelling}
\label{subsec:reg}
\begin{figure}[t]
    \centering
    \includegraphics[width=0.95\textwidth]{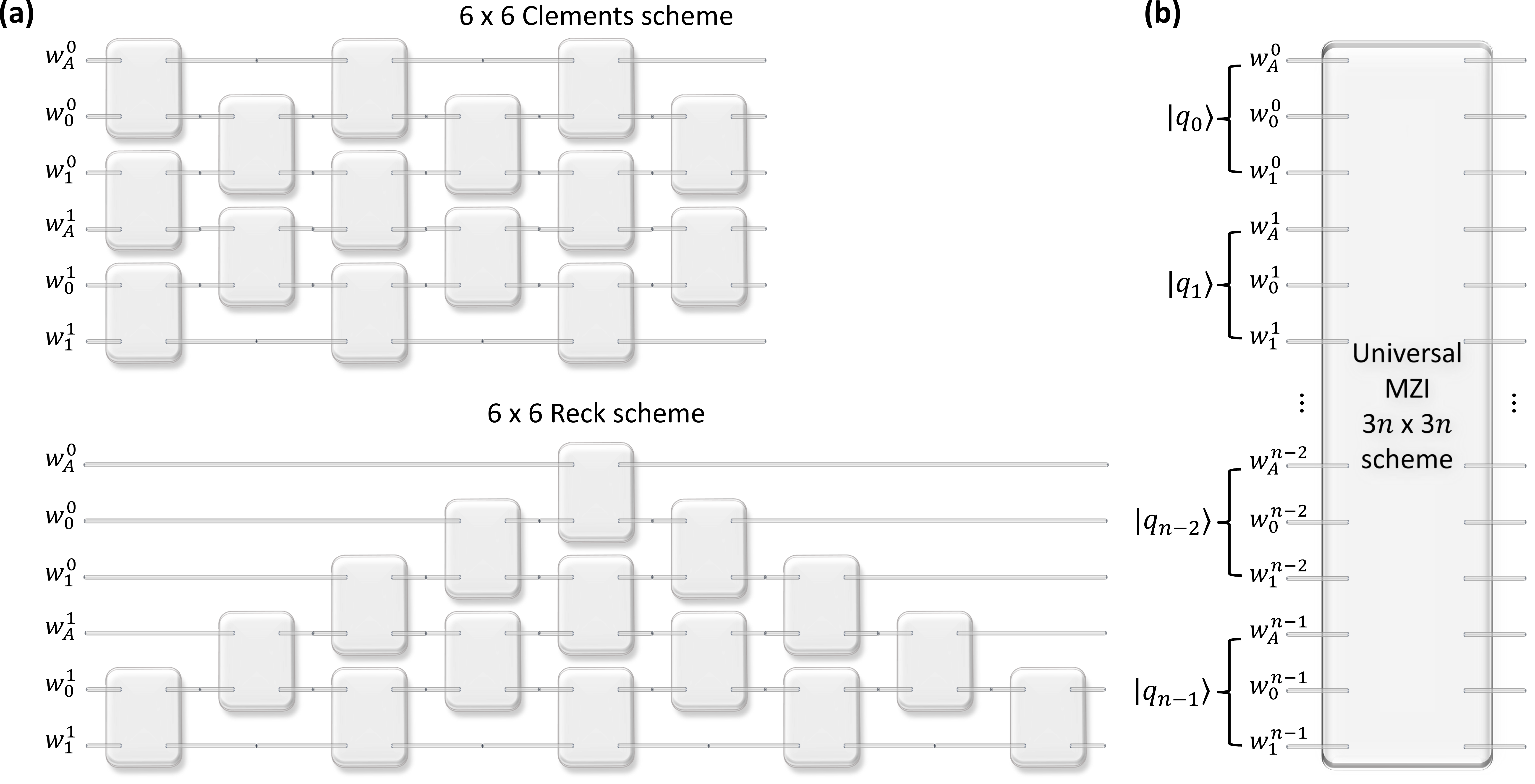}
    \caption{(a) $6\times6$ Clements scheme, upper part, and Reck scheme, lower part, where we embed the regularly-labeled structure of two path-encoded qubits.
    (b) A $3n \times3n$ scheme together with the regular labeling of $n$ path-encoded qubits.}
    \label{fig:unischeme}
\end{figure}
Maintaining a regularly-labeled structure for the path-encoded qubits is a requirement to compute massive quantum algorithms.
The regularly-labeled universal scheme of a MZI network is depicted in Fig.~\ref{fig:unischeme}(b).
The labeling sequentially allocates one ancillary waveguide and a pair of waveguides for path-encoded computational state $\vert \mathsf{0} \rangle, \vert \mathsf{1} \rangle$ so that $(w_A^0, w_0^0, w_1^0)$ corresponds to qubit $|q_0\rangle$, $(w_A^1, w_1^1, w_1^1)$ to $|q_1\rangle$ and so on.
Thus, as described in Sec.~\ref{subsec:qubit}, every path-encoded qubit is composed of one photon and three waveguides.
Once the qubits are initialized, there are two universal schemes of reconfigurable MZI networks that perform quantum operations: the Reck and Clements schemes.
Fig.~\ref{fig:unischeme}(a) reports their $6\times6$ configuration.

Quantum gates in a regularly-labeled structure are equally implemented by MZIs. 
Single-qubit gates can be trivially achieved through a MZI and their action is described by Eq.~\eqref{eq:MZI_matrix}. This operation connects two waveguides represented in the computational basis $|\mathsf{0}\rangle, |\mathsf{1}\rangle$ of a given qubit and leaves untouched the ancillary waveguides.
Two-qubit gates, like the post-selected CZ, operate with a non-regular structure, where ancillary waveguides are located on both sides of the two pairs of path-encoded qubit waveguides. Thus, we have to adapt the post-selected CZ to the chosen regular structure.
In the next subsection, we henceforth present an optical SWAP gate that changes the pathways of photons to make compatible the use of post-selected CZ and the regularly-labeled structure.

\subsubsection{Optical SWAP by Pauli-$X$ gate}
\begin{figure}[t]
    \centering
    \includegraphics[width=0.70\textwidth]{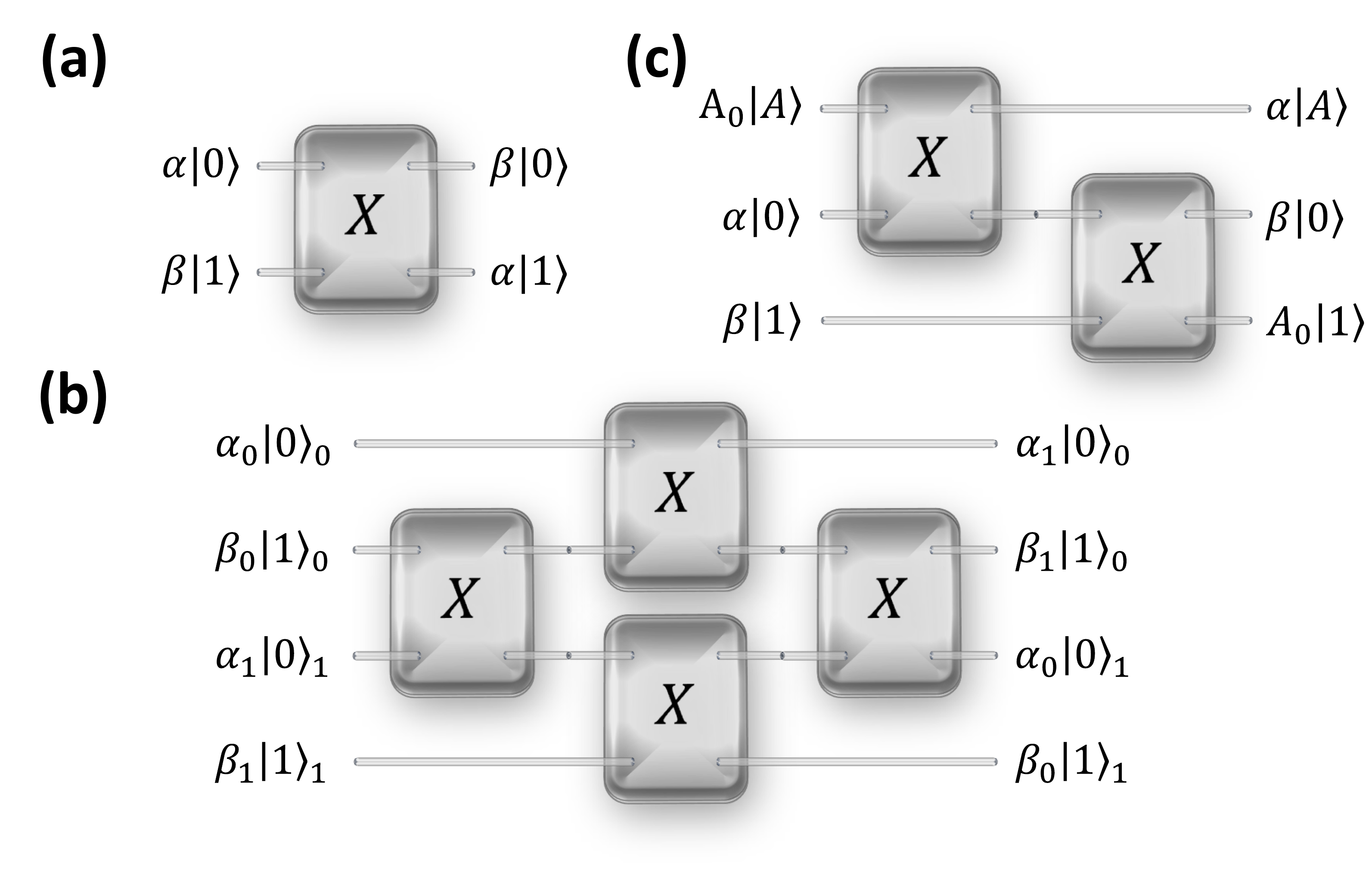}
    \caption{(a) Basic optical SWAP operation on a pair of waveguides. (b) ${\rm SWAP}_2^\prime$ operation which changes the position between two pairs of waveguides. (c) ${\rm SWAP}_1^A$ operation which switches the order between one ancilla waveguide $w_A$ and the pair of waveguides $(w_0,w_1)$. The latter operation is needed to achieve a post-selected CZ gate with a regularly-labeled structure.}
    \label{fig:swap}
\end{figure}
The SWAP operation for a dual-rail qubit is achieved by utilizing a MZI with phases that correspond to a Pauli-$X$ matrix.
Given the generic initial quantum state $\vert \Psi \rangle_{\rm in}$, shown in Eq.~\eqref{eq:generic_1qstate}, we apply the Pauli-$X$ gate
\begin{equation}
\begin{pmatrix}
        \hat{a}_{w_{0}}^\dagger \\
        \hat{a}_{w_{1}}^\dagger 
    \end{pmatrix}
    \to \,\, 
    \begin{pmatrix}
        0 & 1 \\
        1 & 0 
    \end{pmatrix}
    \cdot
    \begin{pmatrix}
        \hat{a}_{w_{0}}^\dagger \\
        \hat{a}_{w_{1}}^\dagger
    \end{pmatrix}\,,
    \label{eq:x_transform}
\end{equation}
and we obtain the output state
\begin{equation}
    \vert \Psi \rangle_{\rm out} = \left( \alpha \,\hat{a}^\dagger_{w_1} + \beta \,\hat{a}^\dagger_{w_0} \right) \vert \Omega \rangle
    = \beta \, \vert \mathsf{0} \rangle + \alpha \, \vert \mathsf{1} \rangle \,.
\end{equation}
We can see that the probability amplitudes of the states are swapped, as depicted in Fig.~\ref{fig:swap}(a). 

Fig.~\ref{fig:swap}(b) shows a two-qubit SWAP gate realized with 4 MZIs, that are configured as Pauli-$X$ gates.
The path-encoded qubits $\vert q_0\rangle$ and $\vert q_1\rangle$ are defined by the position of one photon in each pair of waveguides $(w_0^0, w_1^0)$ and $(w_0^1, w_1^1)$ and, only in this case, we do not consider the auxiliary waveguides.
The transformation given by such a MZI network operates as
\begin{equation}
    \hat{\rm V}_2
    \to
    {\rm SWAP}_2'
    \cdot
    \hat{\rm V}_2 \,,
    \label{eq:oSWAP_two}
\end{equation}
where
\begin{equation}
\begin{split}
    \hat{\rm V}_2^{\rm T} &\equiv 
    \left(
        \hat{a}_{w_{0}^{0}}^\dagger,
        \hat{a}_{w_{1}^{0}}^\dagger,
        \hat{a}_{w_{0}^{1}}^\dagger,
        \hat{a}_{w_{1}^{1}}^\dagger
    \right) \,,\\
{\rm SWAP}_2' &\equiv {\rm X}^{(2,3)}_4\cdot{\rm X}^{(1,2)}_4\cdot{\rm X}^{(3,4)}_4\cdot{\rm X}^{(2,3)}_4 = 
\begin{pmatrix}
        0 & 0 & 1 & 0 \\
        0 & 0 & 0 & 1 \\
        1 & 0 & 0 & 0 \\
        0 & 1 & 0 & 0
    \end{pmatrix} \,,
\end{split}
\label{eq:oSWAP_twobis}
\end{equation}
and ${\rm X}^{(k,k-1)}_m$ represents the single MZI Pauli-$X$ action on the k-th and (k+1)-th waveguides embedded in the $m\times m$ network. In App.~\ref{app:mxm}, embedding a $2\times2$ MZI in a generic $m\times m$ structure is presented in more detail.
Given the initial state as in Eq.~\eqref{eq:generic_2qstate}, the output state reads
\begin{equation}
\begin{split}
    \vert \Psi \rangle_{\rm out} &= \left( \alpha_0 \,\hat{a}^\dagger_{w_0^1} + \beta_0 \,\hat{a}^\dagger_{w_1^1} \right) \left( \alpha_1 \,\hat{a}^\dagger_{w_0^0} + \beta_1 \,\hat{a}^\dagger_{w_1^0} \right) \vert \Omega \rangle 
    = \left( \alpha_1 \,\hat{a}^\dagger_{w_0^0} + \beta_1 \,\hat{a}^\dagger_{w_1^0} \right) \left( \alpha_0 \,\hat{a}^\dagger_{w_0^1} + \beta_0 \,\hat{a}^\dagger_{w_1^1} \right) \vert \Omega \rangle \\
    &= \left( \alpha_{1} \, \vert \mathsf{0} \rangle_0 + \beta_{1} \, \vert \mathsf{1} \rangle_0 \right) \otimes \left( \alpha_{0} \, \vert \mathsf{0} \rangle_1 + \beta_{0} \, \vert \mathsf{1} \rangle_1 \right) \,.
\end{split}
\end{equation}
In other words, the operation swaps the two qubits with respect to the computational basis.
We can note that the transformation is completed only by three layers of MZIs.

Finally, Fig.~\ref{fig:swap}(c) shows the optical SWAP on one photonic qubit composed of the ancillary waveguide and the two waveguides $(w_0,w_1)$.
This process shifts the position of the ancilla waveguide and the quantum states' paths such that
\begin{equation}
    {\rm V}_1
    \to
    {\rm SWAP}_1^A
    \cdot
    {\rm V}_1 \,,
    \label{eq:oSWAP_label}
\end{equation}
where
\begin{equation}
\begin{split}
    {\rm V}_1^{\rm T} &\equiv 
    \left(
        \hat{a}_{w_{A}}^\dagger,
        \hat{a}_{w_{0}}^\dagger,
        \hat{a}_{w_{1}}^\dagger
    \right) \,,\\
{\rm SWAP}_1^A &\equiv {\rm X}^{(0,1)}_3\cdot{\rm X}^{(A,0)}_3 = 
\begin{pmatrix}
        0 & 1 & 0 \\
        0 & 0 & 1 \\
        1 & 0 & 0
    \end{pmatrix} \,.
\end{split}
\end{equation}
This operation is required to implement the post-selected CZ in a regularly-labeled structure as we derive in the next subsection.

\subsection{Requisites on quantum gate level}
\subsubsection{Qubit SWAP by optical SWAP gate}
\begin{figure}[t]
    \centering
    \includegraphics[width=0.70\textwidth]{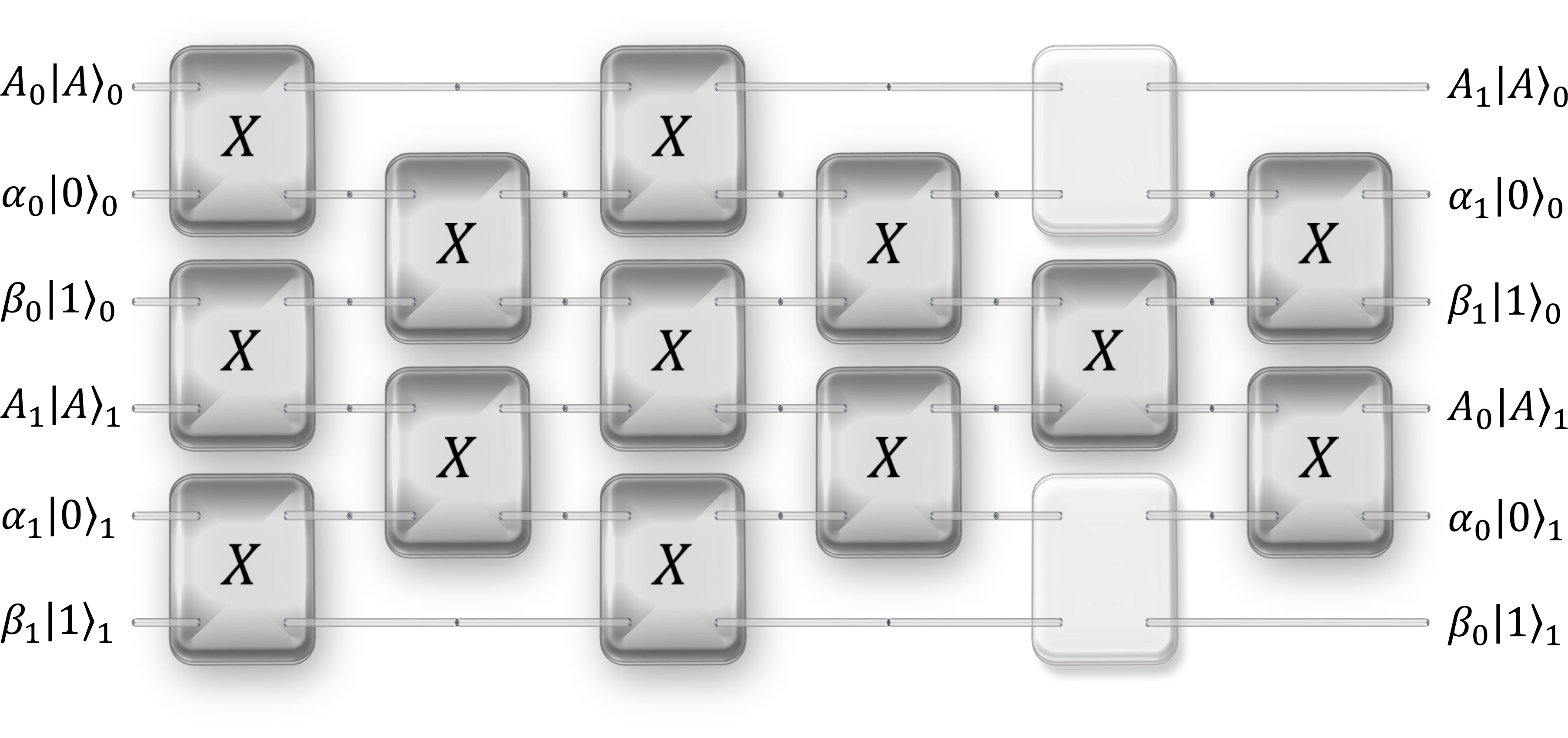}
    \caption{Qubit SWAP operation ${\rm SWAP}_2^A$ in a regularly-labeled structure. This gate exchanges the positions of the ancillary waveguides, i.e. $w_A^0 \leftrightarrow w_A^1$, and the waveguides of the first and second qubits, i.e. $(w_0^0,w_1^0) \leftrightarrow (w_0^1,w_1^1)$.
    The state $\vert A \rangle$ is used to denote the state corresponding to one photon in the auxiliary waveguide, and it is outside the set of states that defines the qubit structures. The MZIs represented by light grey boxes are set in the identity configuration, while the others with darker grey and the letter ''X'' in the Pauli-$X$ setting.}
    \label{fig:qubitswap_1}
\end{figure}
\begin{figure}[t]
    \centering
    \includegraphics[width=0.85\textwidth]{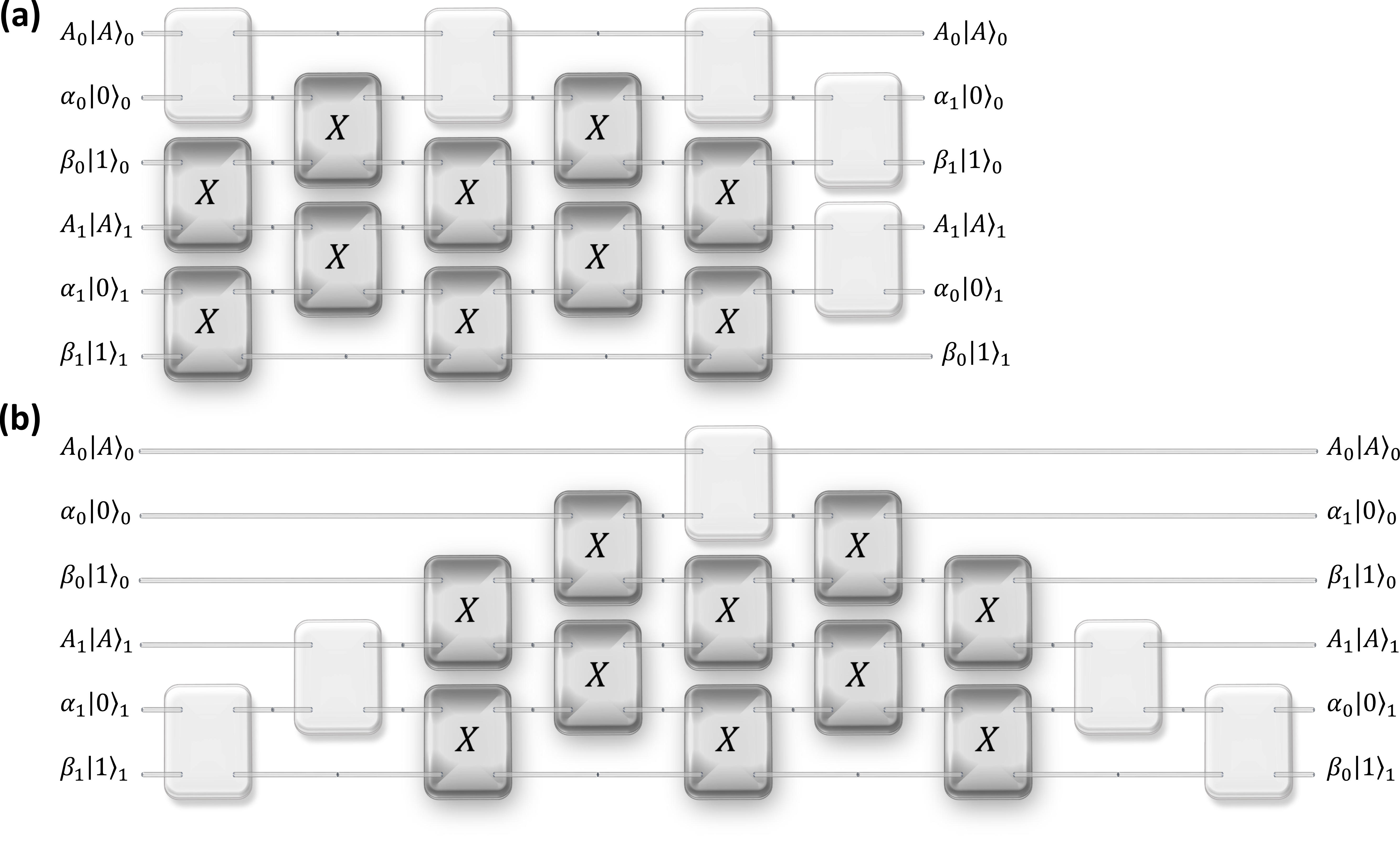}
    \caption{Qubit SWAP operation ${\rm SWAP}_2$ that leaves untouched the ancillary waveguides and exchanges the positions of the waveguides of the first and second qubits, i.e. $(w_0^0,w_1^0) \leftrightarrow (w_0^1,w_1^1)$, in a regularly-labeled structure.
    In particular, we report the embedding of this SWAP gate (a) in 6 $\times$ 6 Clements scheme and (b) in 6 $\times$ 6 Reck scheme.
    The state $\vert A \rangle$ denotes the state corresponding to one photon in the auxiliary waveguide, and it is outside the states that define the qubit structures. The MZIs represented by light grey boxes are set in the identity configuration, while the others with darker grey and the letter ''X'' in the Pauli-$X$ setting.}
    \label{fig:qubitswap_2}
\end{figure}
In this subsection, we perform not only the optical SWAP to switch the location of waveguides, but also a general quantum SWAP operation, which extends this concept.
Two SWAP operations are described in Fig.~\ref{fig:qubitswap_1} and \ref{fig:qubitswap_2}, the former can be embedded in $6\times6$ Clements scheme with regularly-labeled waveguides and the latter can be embedded in both of $6\times6$ Clements and Reck schemes.

The Pauli-$X$ network shown in Fig.~\ref{fig:qubitswap_1} acts as 
\begin{equation}
    {\rm V}_2
    \to
    {\rm SWAP}_2^{A}
    \cdot
    {\rm V}_2 \,,
\end{equation}
where
\begin{equation}
\begin{split}
    {\rm V}_2^{\rm T} &\equiv 
    \left(
        \hat{a}_{w_{A}^{0}}^\dagger,
        \hat{a}_{w_{0}^{0}}^\dagger,
        \hat{a}_{w_{1}^{0}}^\dagger,
        \hat{a}_{w_{A}^{1}}^\dagger,
        \hat{a}_{w_{0}^{1}}^\dagger,
        \hat{a}_{w_{1}^{1}}^\dagger
    \right) \,,\\
    {\rm SWAP}_2^{A} &=
    \mathbb{X}_2 \cdot {\rm X}^{(3,4)}_6 \cdot \mathbb{X}_2 \cdot \mathbb{X}_1^A \cdot  \mathbb{X}_2 \cdot \mathbb{X}_1^A
    =
    \begin{pmatrix}
    0 & 0 & 0 & 1 & 0 & 0 \\
    0 & 0 & 0 & 0 & 1 & 0 \\
    0 & 0 & 0 & 0 & 0 & 1 \\
    1 & 0 & 0 & 0 & 0 & 0 \\
    0 & 1 & 0 & 0 & 0 & 0 \\
    0 & 0 & 1 & 0 & 0 & 0
    \end{pmatrix} \,,\\
    \mathbb{X}_1^A &\equiv {\rm X}^{(1,2)}_6 \cdot {\rm X}^{(3,4)}_6 \cdot {\rm X}^{(5,6)}_6 \,,
    \\
    \mathbb{X}_2 &\equiv {\rm X}^{(2,3)}_6 \cdot {\rm X}^{(4,5)}_6 \,.
\end{split}
\end{equation}

Given the generic initial state of two qubits as in Eq.~\eqref{eq:generic_2qstate}, the action of this transformation consists of the SWAP between qubits in a regularly-labeled structure. 
However, its action is more general than the one reported in Eq.~(\ref{eq:oSWAP_two}-\ref{eq:oSWAP_twobis}) and, since the MZI network shown in Fig.~\ref{fig:qubitswap_1} performs also the swapping between the auxiliary waveguides, or equivalently $(w_A^0 \leftrightarrow w_A^1)$. 
The network of Fig.~\ref{fig:qubitswap_1} is implemented by setting 13 MZIs in the Pauli-$X$ configuration and 2 MZIs as the Identity in the $6\times6$ Clements scheme. 
This network cannot be embedded in the $6\times6$ Reck scheme, since two layers with 3 MZIs are required.

It is important to note that the swapping of auxiliary waveguides is not needed, because the qubit structure selects the events with no photons in the auxiliary waveguides. Indeed, post-selection implies that when a photon remains in any auxiliary waveguide after a two-qubit gate, we do not count the corresponding event on the detectors. 

Therefore, we can optimize the resources as in the Pauli-$X$ network shown in Fig.~\ref{fig:qubitswap_2}, whose action can be written as 
\begin{equation}
    {\rm V}_2
    \to
    {\rm SWAP}_2
    \cdot
    {\rm V}_2 \,,
\end{equation}
where
\begin{equation}
\begin{split}
    {\rm SWAP}_2 &=
    \mathbb{X}_1 \cdot \mathbb{X}_2 \cdot \mathbb{X}_1 \cdot \mathbb{X}_2 \cdot \mathbb{X}_1
    =
    \begin{pmatrix}
    1 & 0 & 0 & 0 & 0 & 0 \\
    0 & 0 & 0 & 0 & 0 & 1 \\
    0 & 0 & 0 & 0 & 1 & 0 \\
    0 & 0 & 0 & 1 & 0 & 0 \\
    0 & 0 & 1 & 0 & 0 & 0 \\
    0 & 1 & 0 & 0 & 0 & 0
    \end{pmatrix} \,, \\
    \mathbb{X}_1 &\equiv {\rm X}^{(3,4)}_6 \cdot {\rm X}^{(5,6)}_6 \,,
    \\
    \mathbb{X}_2 &\equiv {\rm X}^{(2,3)}_6 \cdot {\rm X}^{(4,5)}_6 \,.
    \label{eq:swap2}
\end{split}
\end{equation}
In this way, the action of this transformation involves the SWAP between qubits in a regularly-labeled structure without affecting the auxiliary waveguides. 
This network is composed of 10 MZIs in the Pauli-$X$ configuration and 5 as the Identity gate and it can be embedded in both the $6\times6$ Clements and Reck schemes as shown in Fig.~\ref{fig:qubitswap_2}(a) and (b).

\subsubsection{Two-qubit gate in regularly-labeled waveguides}
\label{subsec:twoop_reg}
\begin{figure}[t]
    \centering
    \includegraphics[width=0.85\textwidth]{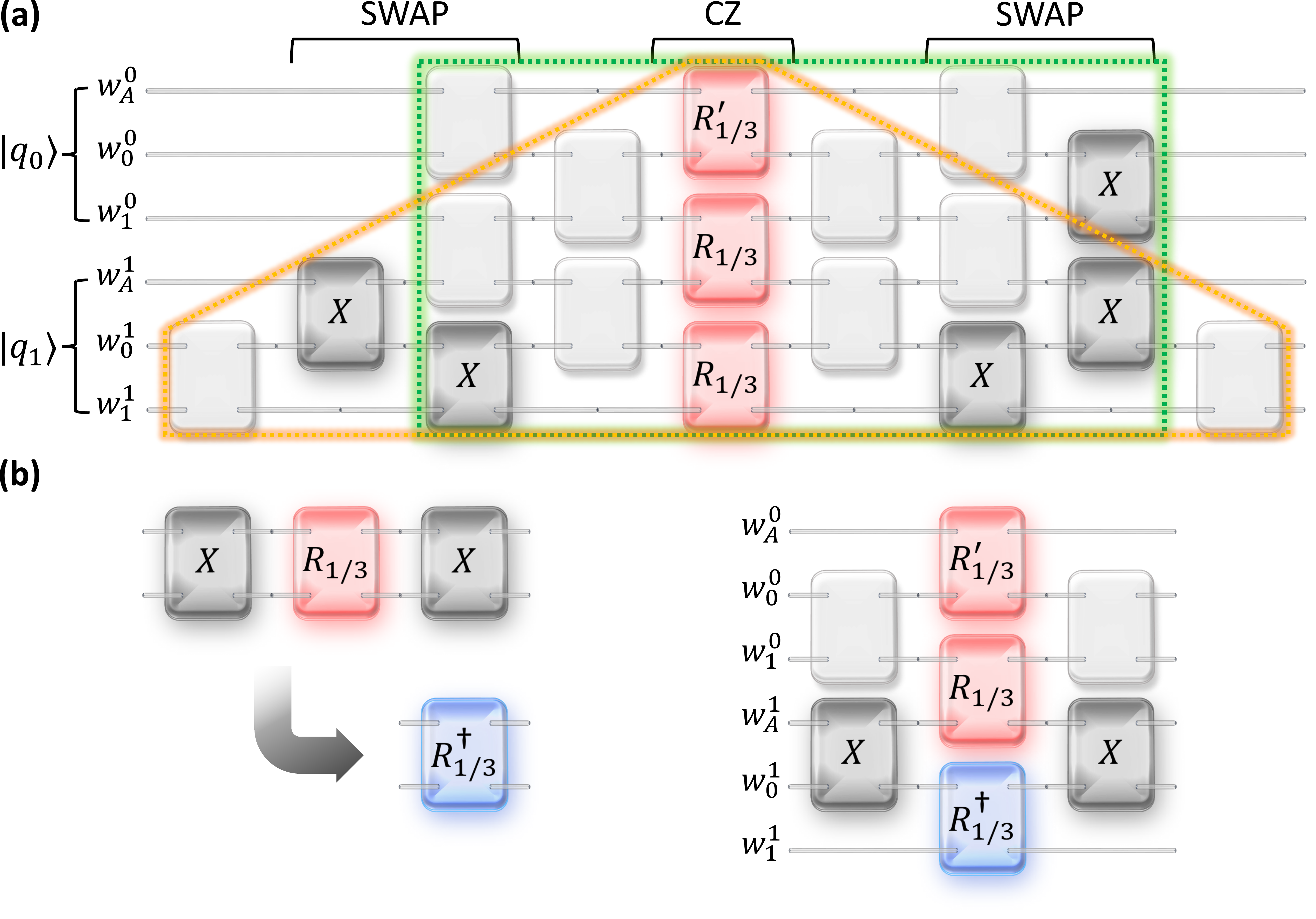}
    \caption{CZ gate implementation in a regularly-labeled structure. 
    (a) Post-selected CZ operation together with optical SWAP that changes the positions of waveguides to map the regular structure to the structure needed for the standard post-selected CZ. Green contours show the $6\times6$ Clements scheme and orange contours the $6\times6$ Reck scheme.
    (b) Compressed CZ implementation in the regularly-labeled structure.
    The MZIs represented with light grey boxes are set in the identity configuration, while the colored ones are set in the configurations X, $R_{1/3}$, $R'_{1/3}$ and $R^\dagger_{1/3}$, marked with the corresponding label.}
    \label{fig:compress_cz}
\end{figure}
In this subsection, we derive the regularly-labeled post-selected CZ by restoring the labeling of waveguides using the optical SWAP proposed in the previous subsection.

Firstly, we consider the initial state of two qubits given in Eq.~\eqref{eq:generic_2qstate}, and we apply the optical SWAP to the target qubit. 
The state evolves through the transformation given in Eq.~\eqref{eq:oSWAP_label} as 
\begin{equation}
\begin{aligned}
    \vert \Psi \rangle_{\rm in} & = \vert q_0 \rangle_{\rm in} \otimes \vert q_1 \rangle_{\rm in} \\
    &\rightarrow 
    \left( \mathbf{1}\otimes{\rm SWAP}_1^A \right)\,
    \vert q_0 \rangle_{\rm in} \otimes \vert q_1 \rangle_{\rm in} \,.
\end{aligned}
\end{equation} 
The previous transformation maps the regularly-labeled pathway of photons into the positions required by the post-selected CZ.
Then, we act with the CZ, as it is explained in Sec.~\ref{subsec:twoop}, and lastly we again apply the optical SWAP symmetrically. The final state reads
\begin{equation}
\begin{aligned}
    &\left( \mathbf{1}\otimes\left({\rm SWAP}_1^A\right)^{\rm T} \right)
    \cdot
    \overline{\rm CZ}_{\rm ps} \cdot
    \left( \mathbf{1}\otimes{\rm SWAP}_1^A \right) \,
    \vert q_0 \rangle_{\rm in} \otimes \vert q_1 \rangle_{\rm in} \\
    &= -\frac{1}{3} \left( \alpha_{0} \alpha_{1} \,\hat{a}_{w_{0}^{0}}^\dagger \hat{a}_{w_{0}^{1}}^\dagger + \alpha_{0}\beta_{1}\,\hat{a}_{w_{0}^{0}}^\dagger\hat{a}_{w_{1}^{1}}^\dagger 
        + \beta_{0} \alpha_{1}\,\hat{a}_{w_{1}^{0}}^\dagger \hat{a}_{w_{0}^{1}}^\dagger - \beta_{0} \beta_{1} \,\hat{a}_{w_{1}^{0}}^\dagger \hat{a}_{w_{1}^{1}}^\dagger\right) \vert \Omega \rangle 
        +\ldots \\
        & \implies \alpha_{0} \alpha_{1} \,\vert \mathsf{00} \rangle + \alpha_{0} \beta_{1} \,\vert \mathsf{01} \rangle  
        + \beta_{0} \alpha_{1} \,\vert \mathsf{10} \rangle - \beta_{0} \beta_{1} \,\vert \mathsf{11} \rangle  
    \,.
\end{aligned}
\label{eq:cz_substitution_reg}
\end{equation} 
This result is exactly the one reported in Eqs.~\eqref{eq:cz_hp} and \eqref{eq:cz_corr}.
This gate can be realized only in the $6\times6$ Reck scheme as it is shown in Fig.~\ref{fig:compress_cz}(a).

However, we can shrink the full depth of post-selected CZ.
To do this, we note that the bottom $R_{1/3}^{\dagger}$ of the post-selected CZ is located between two Pauli-$X$ MZI. Thus, we can do the three operations simultaneously by introducing the new rotation operator $R_{1/3}^{\dagger}$ to compress the regularly-labeled CZ gate as depicted in Fig.~\ref{fig:compress_cz}(b).
The operator can be derived by multiplying Pauli-$X$ on each side of $R_{1/3}$
\begin{equation}
    R_{1/3}^{\dagger} \equiv
    {\rm X} \cdot R_{1/3} \cdot {\rm X} = 
    \frac{1}{\sqrt{3}}
    \begin{pmatrix}
        -1 & -\sqrt{2} \\
        \sqrt{2} & -1
    \end{pmatrix} \,,
\end{equation}
and it is achieved by setting $\theta_1-\theta_2 = - 2 \arcsin \frac{1}{\sqrt{3}}$ and $\phi_1-\phi_2 =0$.
This new gate replaces three MZIs, since no quantum gate is applied between CZ and optical SWAP.

Finally, we consider the generic state given in Eq.~\eqref{eq:oSWAP_label} and the action of the regularly-labeled CZ, which reads
\begin{equation}
    {\rm V}_2
    \to
    {\rm CZ}_{\rm ps}
    \cdot
    {\rm V}_2 \,,
\end{equation}
where
\begin{equation}
\begin{split}
{\rm CZ}_{\rm ps} &\equiv  {\rm X}^{(w_A^1,w_0^1)}_6 \cdot \overline{\rm CZ}_{\rm ps} \cdot {\rm X}^{(w_A^1,w_0^1)}_6 
=
\frac{1}{\sqrt{3}}
\footnotesize
\begin{pmatrix}
    -1 & \sqrt{2} &  0 & 0 & 0 & 0 \\
    \sqrt{2} & 1 & 0  & 0 & 0 & 0 \\
    0 & 0 & -1 & 0 & -\sqrt{2} & 0 \\
    0 & 0 & 0 & -1 & 0 & \sqrt{2} \\
    0 & 0 & \sqrt{2} & 0 & -1 & 0 \\
    0 & 0 & 0 & -\sqrt{2} & 0 & -1 \\
    \end{pmatrix}\,.
\end{split}
\end{equation}
It is straightforward to check that the transformation
\begin{equation}
\begin{aligned}
    \vert \Psi \rangle_{\rm in} & = \vert q_0 \rangle_{\rm in} \otimes \vert q_1 \rangle_{\rm in} \\
    &\rightarrow 
    {\rm CZ}_{\rm ps}\,
    \vert q_0 \rangle_{\rm in} \otimes \vert q_1 \rangle_{\rm in} 
\end{aligned}
\end{equation} 
gives the same result of Eqs.~\eqref{eq:cz_hp}, \eqref{eq:cz_corr} and also \eqref{eq:cz_substitution_reg}.
From Fig.~\ref{fig:compress_cz}(b), we can observe that the structure of the new post-selected CZ, also called regularly-labeled post-selected CZ, can be easily embedded in the $6\times6$ universal schemes. 
In particular, there is one way of embedding it in the Reck scheme and two ways in the Clements one.

We conclude this section by showing that a regularly-labeled CNOT can be realized as 
\begin{equation}
    {\rm CNOT}_{\rm ps} = \left( \mathbf{1}\otimes {\rm H} \right) \cdot {\rm CZ}_{\rm ps} \cdot \left( \mathbf{1}\otimes {\rm H} \right) \,,
    \label{eq:cnot_compressed}
\end{equation}
where the transformation $\mathbf{1}\otimes {\rm H}$ represents the action of H gate on the target qubit.

\section{Bell state and GHZ state}
\label{sec:example}
\begin{figure}[t]
    \centering
    \includegraphics[width=0.9\textwidth]{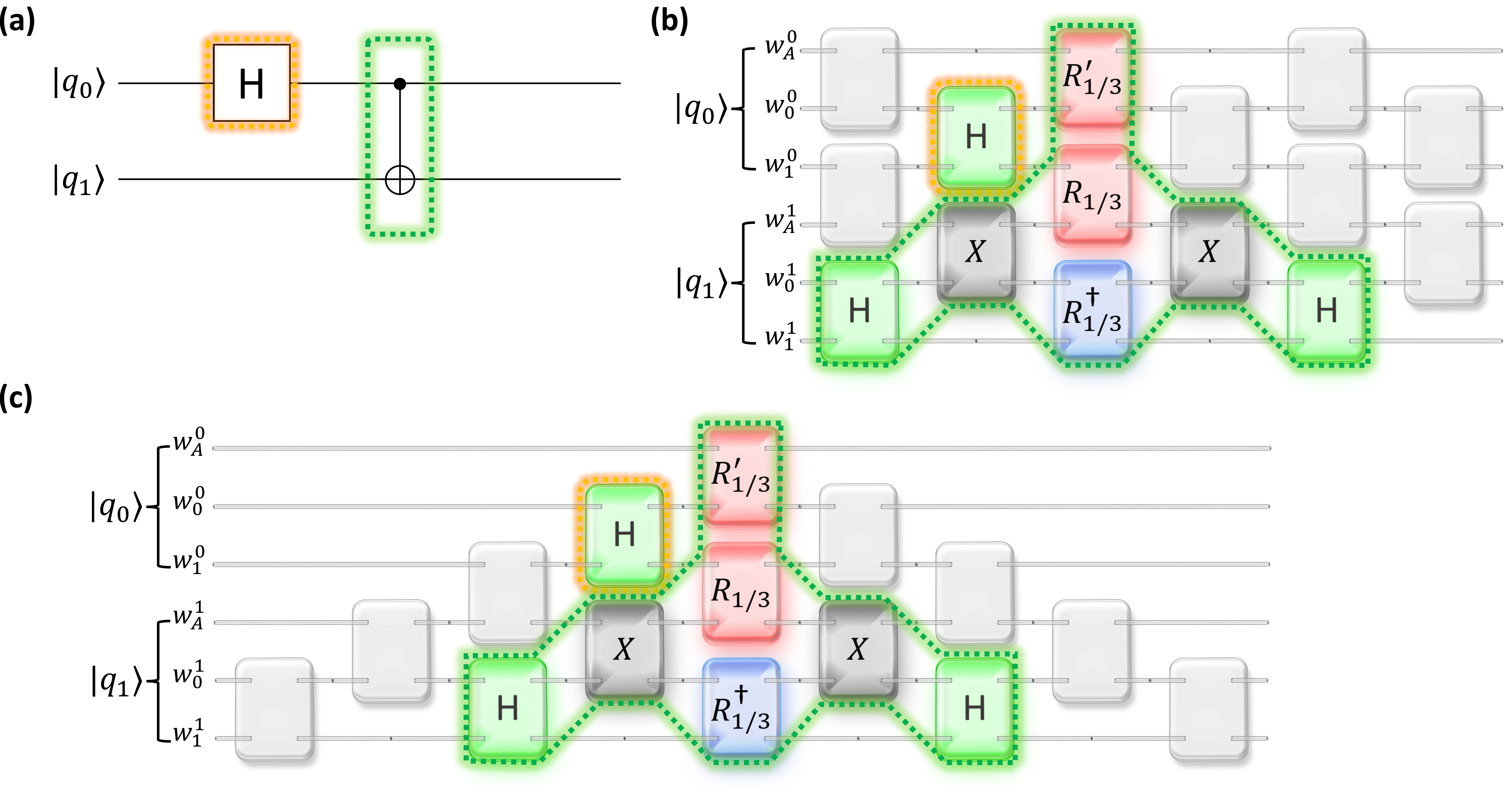}
    \caption{Bell state generation in both quantum circuit representation and MZI network.
    (a) Gate-based Bell state generation circuit.
    The corresponding embedding of Bell state generation circuit (b) in the $6\times6$ Clements scheme and (c) in the $6\times6$ Reck scheme. 
    The MZIs represented with light grey boxes are set in the identity configuration, while the colored ones are set in the configurations X, H and the different $1/3$-BSs.
    The yellow contour enclosing the MZI, which is set in the H configuration, corresponds to the H gate of the gate-based Bell state generation circuit. Analogously, the green contour enclosing the MZI network corresponds to the CNOT gate of the gate-based Bell state generation circuit.
    }
    \label{fig:bellstate}
\end{figure}
This section shows examples of quantum circuits based on our proposed regularly-labeled structure.
We present the circuits to generate the Bell and the GHZ states.
In both examples, all the qubits are initially prepared in the $|\mathsf{0}\rangle$ state.

The Bell state generation is described in Fig.~\ref{fig:bellstate}(a): the circuit is composed of one H gate that is applied to the first qubit $\vert q_0 \rangle$ and a CNOT gate bonding the pair $(\vert q_0\rangle, \vert q_1 \rangle)$, where the first qubit acts as the control and the second as the target.
We propose to implement this circuit by utilizing the path-encoded qubits made of bundles of three waveguides and a $6\times6$ regularly-labeled MZI network.
Thus, we assign the qubits $\vert q_0\rangle$ and $\vert q_1\rangle$ to the triplets of waveguides $(w_A^0, w_0^0, w_1^0)$ and $(w_A^1, w_0^1, w_1^1)$, respectively.
Quantum gates are sequentially mapped into proper locations of MZI, by using the notion of Sec.~\ref{subsec:singleop} for the single-qubit gates and the regularly-labeled compressed post-selected CNOT, discussed in Sec.~\ref{subsec:twoop_reg} and illustrated in Fig.~\ref{fig:compress_cz}.
Fig.~\ref{fig:bellstate}(b--c) shows the embedding of the Bell state generation circuit in $6\times6$ universal MZI networks.
More explicitly, given the initial state $\vert \mathsf{00} \rangle$, the scheme is described by the following transformation:
\begin{equation}
\begin{split}
     \vert \mathsf{00} \rangle
     \to
     {\rm CNOT}_{\rm ps} \cdot \left(  {\rm H} \otimes \mathbf{1}\right)  
     \, \vert \mathsf{00} \rangle = \frac{1}{\sqrt{2}} \left( \vert \mathsf{00} \rangle + \vert \mathsf{11} \rangle \right) \,,
     \end{split}
\end{equation}
where we use the regularly-labeled compressed post-selected CNOT given in Eq.~\eqref{eq:cnot_compressed}.
We point out that the post-selection is required and the success probability of this Bell state generation circuit is $1/9$.

\begin{figure*}[t]
    \centering
    \includegraphics[width=0.95\textwidth]{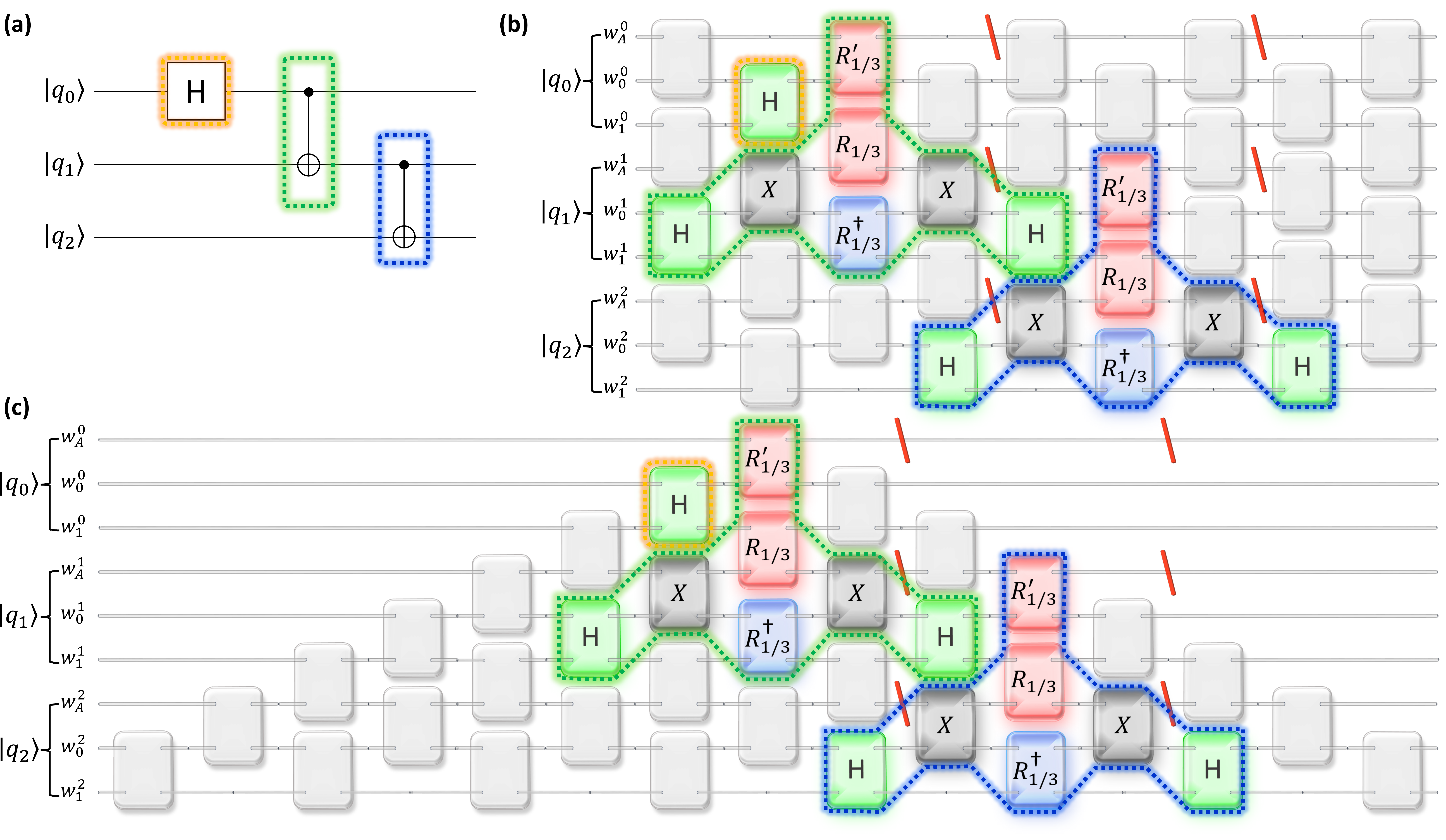}
    \caption{GHZ state generation in a first example of quantum circuit representation and corresponding MZI networks.
    (a) A first example of gate-based GHZ state generation circuit.
    The corresponding GHZ state generation circuit embedded in universal $9\times9$ (b) Clements and (c) Reck MZI networks.
    The MZIs represented with light grey boxes are set in the identity configuration, while the colored ones are set in the configurations X, H and the different $1/3$-BSs.
    The red diagonal lines on auxiliary waveguides represent the {\it truncation trick}: in such positions the waveguides are not connected.
    The yellow contour enclosing the MZI, which is set in the H configuration, corresponds to the H gate of the gate-based GHZ state generation circuit. Analogously, the green and blue contours enclosing the MZI networks correspond to the first and second CNOT gates of the gate-based GHZ state generation circuit.
    }
    \label{fig:ghzstate}
\end{figure*}
\begin{figure*}[t]
    \centering
    \includegraphics[width=0.95\textwidth]{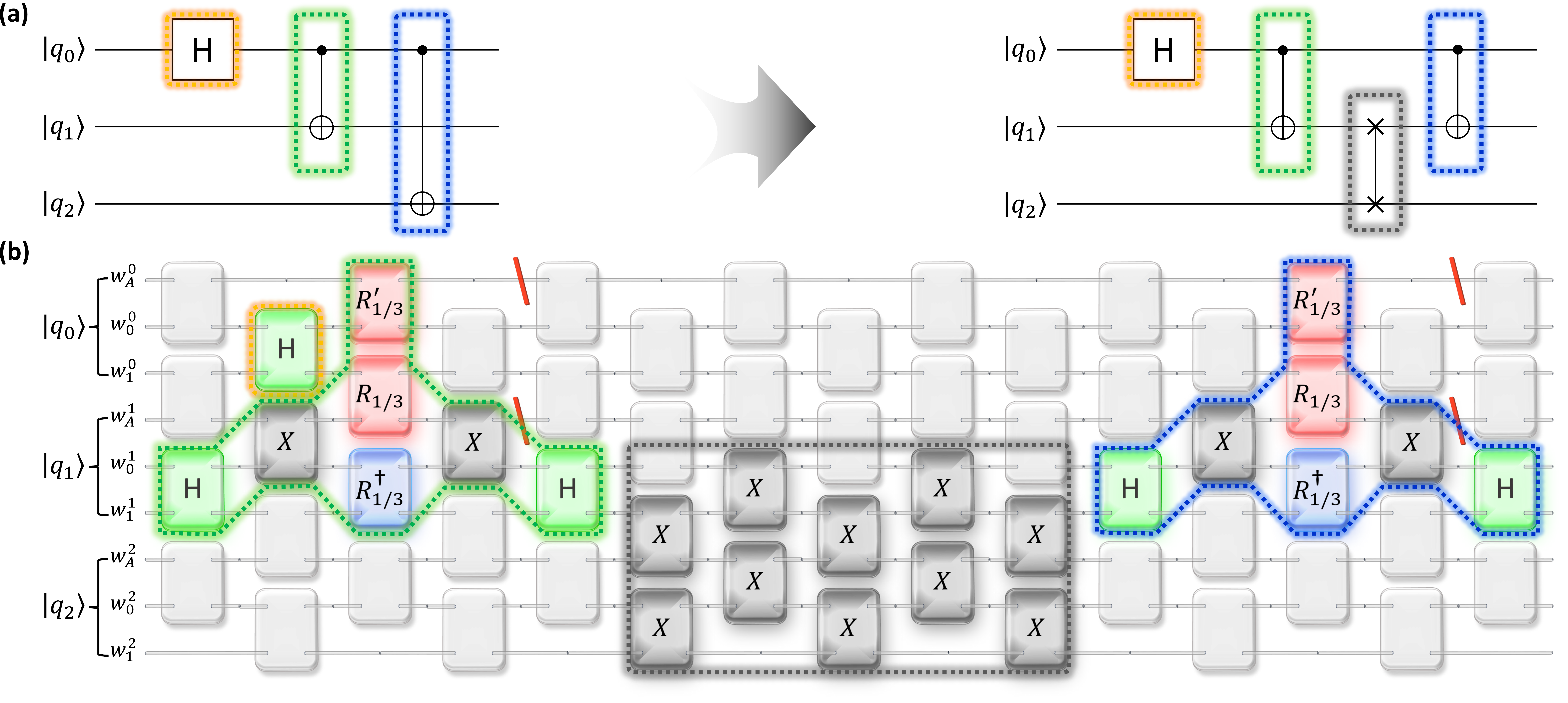}
    \caption{GHZ state generation in a second example of quantum circuit representation and corresponding MZI networks.
    (a) A second example of gate-based GHZ state generation circuit, that involves the use of the SWAP gate.
    (b) The corresponding GHZ state generation circuit, embedded in a MZI network.
    The MZIs represented with light grey boxes are set in the identity configuration, while the colored ones are set in the configurations X, H and the different $1/3$-BSs.
    The red diagonal lines on auxiliary waveguides represent the {\it truncation trick}: in such positions the waveguides are not connected.
    The yellow contour enclosing the MZI, which is set in the H configuration, corresponds to the H gate of the gate-based GHZ state generation circuit. Analogously, the green and blue contours enclosing the MZI networks correspond to the first and second CNOT gates of the gate-based GHZ state generation circuit.
    Finally, the grey contour stands for the SWAP gate implemented by $6\times6$ Clements scheme to connect non-nearest neighbour path-encoded qubits.
    }
    \label{fig:ghzstate_swap}
\end{figure*}

Next, GHZ state generation circuits, whose gate-based circuit is shown in Fig.~\ref{fig:ghzstate}(a), are shown as an example of computational scalability in photonic integrated circuits.
We use the same strategy as in the previous example, so we map the qubits $\vert q_0\rangle$, $\vert q_1\rangle$ and $\vert q_2\rangle$ to the three bundles of waveguides, $(w_A^0, w_0^0, w_1^0)$, $(w_A^1, w_0^1, w_1^1)$, $(w_A^2, w_0^2, w_1^2)$, respectively. Then, we consider $9\times9$ universal MZI networks.
The circuit involves an additional CNOT gate to the pairs $(\vert q_1\rangle, \vert q_2\rangle)$ or $(\vert q_0\rangle, \vert q_2\rangle)$ compared to the previous case, as depicted in Fig.~\ref{fig:ghzstate}.
The first case is shown in Fig.~\ref{fig:ghzstate}(b--c): we can note that it is possible to embed such GHZ state generation circuit in $9\times9$ universal MZI networks.
The second case is illustrated in Fig.~\ref{fig:ghzstate_swap} and it shows an example of SWAP operation between qubits, described in Eq.~\eqref{eq:swap2} and represented in Fig.~\ref{fig:qubitswap_2}.
In this case, applying the SWAP operation is needed because the physical implementation of CNOT does not allow for operations between non-nearest-neighbor path-encoded qubits.
In both cases, the {\it truncation trick} is required since it allows to cascade the post-selected CZ in the utilized configurations.
Again, given the initial state $\vert \mathsf{000} \rangle$, the schemes in Fig.~\ref{fig:ghzstate}--\ref{fig:ghzstate_swap} are described by the following transformations:
\begin{equation}
\begin{split}
     & \vert \mathsf{000} \rangle
     \to U_{\rm GHZ} \, \vert \mathsf{000} \rangle = 
     \frac{1}{\sqrt{2}} \left( \vert \mathsf{000} \rangle + \vert \mathsf{111} \rangle \right) \,, \\
     & U_{\rm GHZ} =
     \hat{\rm P}_{\rm aux}\!
    \cdot\!
    \left(\mathbf{1}\otimes {\rm CNOT}_{\rm ps}^{(2,3)}  \right)\!
    \cdot \hat{\rm P}_{\rm aux}\!
    \cdot \!\left( {\rm CNOT}_{\rm ps}^{(1,2)}\! \otimes \mathbf{1}\right)\!\cdot\! \left(  {\rm H} \otimes \mathbf{1}\otimes \mathbf{1}\right) \,, \\
     & U_{\rm GHZ}^{\rm (swap)} =
     \hat{\rm P}_{\rm aux}\!
    \cdot\!
    \left( {\rm CNOT}_{\rm ps}^{(1,2)} \!\otimes \mathbf{1}\right)
    \!\cdot\!
    \left( \mathbf{1}\otimes {\rm SWAP}_2^{(2,3)}\right) \!
    \cdot \hat{\rm P}_{\rm aux}\!
    \cdot \!\left( {\rm CNOT}_{\rm ps}^{(1,2)} \!\otimes \mathbf{1}\right)\!\cdot \!\left(  {\rm H} \otimes \mathbf{1}\otimes \mathbf{1}\right) \,, \\
\end{split}
\end{equation}
where $\hat{\rm P}_{\rm aux}$ represents the action of the {\it truncation trick} and the upper indexes in parenthesis denote the qubits involved in the operation. We utilize the regularly-labeled compressed post-selected CNOT given in Eq.~\eqref{eq:cnot_compressed} and the optical SWAP given Eq.~\eqref{eq:swap2}.
We point out that the post-selection together with the {\it truncation trick} is required and the success probability of this circuit is $1/81$.
Moreover, in the circuit reported in Fig.~\ref{fig:ghzstate}(b-c)-\ref{fig:ghzstate_swap}(b) we note that the truncation is necessary only on the waveguide $w_A^1$ between the two CNOT gates. We insert all the truncation after any two-qubit gate, since we follow the prescription of the {\it truncation trick} and we allow the possibility of different compositions of CNOT gates between the three qubits.

A generic $n$-dimensional GHZ state,
\begin{equation}
\begin{split}
    \frac{1}{\sqrt{2}} \left( \bigotimes_{j=0}^{n-1} \vert \mathsf{0} \rangle_j + \bigotimes_{j=0}^{n-1} \vert \mathsf{1} \rangle_j \right)
    \,,
\end{split}
\end{equation}
can be obtained by iterating the previous circuits. Its success probability would be $9^{1-n}$, for $n$ bigger than 2.

We conclude this section by emphasizing that the {\it truncation trick}, also shown in Fig.~\ref{fig:ghzstate}, is compulsory in order to cascade the post-selected CZ between different pairs of qubits that share only one qubit. Indeed, without this procedure we do not achieve the correct result for the GHZ state generation circuit, as shown in App.~\ref{app:GHZ}. 
The {\it truncation trick} makes possible the cascading of the post-selected CZ on the pairs of qubits that share only one qubit, but we cannot cascade it on the same pair of qubits twice. 

\section{Conclusion}
\label{sec:conclusion}
Our work presents a scalable possibility of integrated photonic circuits for LOQC.
Regularly-labeled structure for path-encoded qubits and optical SWAP operations are needed. Based on these primitives, we demonstrate qubit SWAP gates to enable universal programming and post-selected CZ gates.
The compressed version of regularly-labeled post-selected CZ is also presented to decrease the depth of the MZI network and to minimize quantum resource consumption.
Finally, we propose a new way of using the post-selected CZ: indeed, through the {\it truncation trick} this gate can be sequentially applied on two pairs of qubits that share only one qubit.
Formally, we present two exemplary quantum circuits: the Bell and the GHZ state generation circuits.

Our approach is still characterized by some problems that must be mitigated in order to reach large-scale computation.
First of all, the success probability scales with multiples of $1/9$ every time a post-selected CZ is used. Moreover, such a gate cannot be cascaded twice to the same pair of qubits, even if we apply the {\it truncation trick}. This implies the impossibility of generating the most general multipartite entangled state.
From this perspective, the measurement-based approach~\cite{bartolucci_fusion-based_2023,briegel2001persistent,raussendorf2001one} through fusion gates has better performances on the integrated photonic platform.

\section*{Acknowlegments}
\label{acknowlegments}
We would like to thank KangHyeon Kim (Pukyong National University) for discussing our main proposals;
Y.~Kwon and B.-S.~Choi were supported in part by the National Research Foundation of Korea (NRF) grant funded by the Korean Government Ministry of Science and ICT (MSIT) under Grant 2020K1A3A1A78087782, and in part by the Pukyong National University Industry-university Cooperation Research Fund in 2023 (202312370001).
The work of UNITN was supported by the Horizon 2020 Framework Programme (899368) and by the Provincia Autonoma di Trento through the Q@TN joint laboratory.

\bibliography{uploqc}

\appendix
\section{The impossibility of CZ gate through a universal MZI network $4\times4$ }
\label{app:4x4imp}

In this section we want to explore the possibility of achieving a CZ gate in path-encoding by using a universal MZI network $4\times4$.

As explained in the main text, we assign to every pair of waveguides the states of the computational basis, Eq.~\eqref{eq:qubit_assigned}. 
Then we consider the transformation of a universal MZI network $4\times4$, which is introduced in the previous section:
\begin{equation}
    \begin{aligned}
    U_{4} =&   
    \begin{pmatrix}
        u_{11} & u_{12} & u_{13} & u_{14} \\
        u_{21} & u_{22} & u_{23} & u_{24}  \\
        u_{31} & u_{32} & u_{33} & u_{34}  \\
        u_{41} & u_{42} & u_{43} & u_{44} 
    \end{pmatrix} 
    \end{aligned} \,.
\end{equation}

As a first step we consider the generic initial state of two qubits
\begin{equation}
\begin{split}
    \vert \Psi_2 \rangle_{\rm in} &= \left( A_0 \,\hat{a}_{0}^\dagger + A_1\, \hat{a}_{1}^\dagger \right)
    \left( B_0\, \hat{b}_{0}^\dagger + B_1\, \hat{b}_{1}^\dagger \right) \vert \Omega \rangle \,,
\end{split}
\end{equation}
we apply the transformation
\begin{equation}
\left(a^\dagger_{0},a^\dagger_{1},b^\dagger_{0},b^\dagger_{1}\right)^{\rm T}
\to U^{-1}_4 \cdot
\left(a^\dagger_{0},a^\dagger_{1},b^\dagger_{0},b^\dagger_{1}\right)^{\rm T} \,,
\end{equation}
to the initial state $\vert \Psi_2 \rangle_{\rm in}$ and we require the following output
\begin{equation}
\begin{split}
    \vert \Psi_2 \rangle_{\rm out} &= c \left( A_0 B_0 \,\hat{a}_{0}^\dagger \hat{b}_{0}^\dagger + A_1B_0\, \hat{a}_{1}^\dagger \hat{b}_{0}^\dagger 
    +A_0 B_1 \,\hat{a}_{0}^\dagger \hat{b}_{1}^\dagger - A_1B_1\, \hat{a}_{1}^\dagger \hat{b}_{1}^\dagger\right) \vert \Omega \rangle +\ldots \,,
\end{split}
\end{equation}
where $c$ is a constant and the dots stand for terms that do not satisfy the qubit structure. Note that $|c|^2$ is the success probability of the gate.

Solving the constraints coming from the desired output we have that
\begin{equation}
    \begin{aligned}
    U^{-1}_{4} =&   
    \begin{pmatrix}
        \gamma_{11} & 0 & 0 & 0 \\
        0 & -\gamma_{11} & \gamma_{23} & 0  \\
        0 & \frac{2c}{\gamma_{23}} & \frac{c}{\gamma_{11}} & 0  \\
        0 & 0 & 0 & \frac{c}{\gamma_{11}}
    \end{pmatrix} 
    \end{aligned} \,,
\end{equation}
where $\gamma$s parameters and $c$ are free parameters.

Finally we require the unitarity of $U_4$, or equivalently $U_4 \cdot U_4^{-1} = \mathbf{1}$, which implies the conservation of probability.
From these conditions we find out that the previous matrix cannot be unitary and there is an inconsistency between the unitary conditions and the requirements on the desired output.
Therefore, it is impossible to construct a post-selected CZ through a universal MZI network $4\times4$.

The same result can be derived for a post-selected version of the CNOT gate in a universal MZI network $4\times4$.

\section{How to relate inputs and outputs of optical devices}
\label{app:mxm}
A device with $m$ inputs and $m$ can be described by a $m\times m$ matrix, 
\begin{equation}
    \begin{aligned}
    U =&   
    \begin{pmatrix}
        u_{11} & \ldots & u_{1m} \\
         \vdots &  \ddots &  \vdots \\
        u_{m1} & \ldots & u_{mm} \\
    \end{pmatrix} 
    \end{aligned}
    \label{unitarygen}
\end{equation}
where the components are complex number and $|\det(U)|=1$, if we have conservation of energy, or probability.

Using the second quantization notation with creation and annihilation operators, we can define the input vector as $(a_1^\dagger,\ldots,a_m^\dagger)$ and the output vectors as $(b_1^\dagger,\ldots,b_m^\dagger)$, where we choose to count the modes from up position to down position.

The matrix $U$ establishes the relations between the inputs and outputs
\begin{align}
    \begin{pmatrix}
        b_1^\dagger \\
        \vdots \\
        b_m^\dagger
    \end{pmatrix}
    =
    \begin{pmatrix}
        u_{11} & \ldots & u_{1m} \\
         \vdots &  \ddots &  \vdots \\
        u_{m1} & \ldots & u_{mm} \\
    \end{pmatrix} 
    \begin{pmatrix}
        a_1^\dagger \\
        \vdots \\
        a_m^\dagger
    \end{pmatrix}
    = \begin{pmatrix}
        \sum_{k=1}^m u_{1k} \,a_k^\dagger \\
        \vdots \\
        \sum_{k=1}^m u_{mk} \,a_k^\dagger
    \end{pmatrix}
    \,.
\end{align}
Inverting the matrix $U$, we can invert the previous equation and use the relation from $a^\dagger$s to $b^\dagger$s to analyse the evolution of a given state.

Let's make the example of $2\times2$ devices. In this case, we simply have 
\begin{align}
    \begin{pmatrix}
        a_1^\dagger \\
        a_2^\dagger
    \end{pmatrix}
    = \frac{1}{\det(U)}
    \begin{pmatrix}
        u_{22} & -u_{12} \\
        -u_{21} & u_{11}
    \end{pmatrix}
    \begin{pmatrix}
        b_1^\dagger \\
        b_2^\dagger
    \end{pmatrix}
    = \frac{1}{\det(U)}
    \begin{pmatrix}
        u_{22} \,b_1^\dagger - u_{12}\,b_2^\dagger \\
        -u_{21} \,b_1^\dagger + u_{11}\,b_2^\dagger
    \end{pmatrix}\,.
    \label{eq:2x2gen}
\end{align}

We conclude this subsection by constructing a generic matrix $m\times m$ from $2\times2$ sub-matrices. Concretely, this is equivalent to creating a generic $m\times m$ device assembling $2\times2$ devices.
In order to do this we define a generic embedded $2\times2$ matrix $U$ in $m\times m$ system. The transformation of the $k$-th and $(k+1)$-th inputs reads 
\begin{equation}
U^{(k,k+1)} \equiv
    \begin{pmatrix}
    1 & 0 &  \ldots & \ldots & \ldots & 0 \\
    0 & \ddots & & & &\vdots \\
    \vdots & & u_{11} & u_{12} & & \vdots \\
    \vdots & & u_{21} & u_{22} & & \vdots \\
    \vdots & & & & \ddots & 0 \\
    0 & \ldots & \ldots & \ldots & 0 & 1 \\
    \end{pmatrix} \,.
    \label{Ukemb}
\end{equation}
This transformation is leaving unaffected all inputs different from $k$ and $(k+1)$ ones, which are evolving accordingly to Eq.~\eqref{eq:2x2gen}.

The Reck and Clements~\cite{reck_experimental_1994,clements_optimal_2016} schemes give two prescriptions to obtain a generic unitary operation, described by the transformation~\eqref{unitarygen}, by assembling the MZIs, or equivalently the transformations~\eqref{Ukemb}.
These procedures are used for $m\times m$ networks of MZI.

\section{How to cascade post-selected CZ }
\label{app:GHZ}

Let's consider six waveguides, where we assign the first qubit to the first three waveguides and the second qubit to the latter waveguides.
We can start by putting all the creation operators in a vector as
\begin{equation}
    \overline{\rm V}_2^{\rm T} \equiv \left((a^\dagger_{A},a^\dagger_{0},a^\dagger_{1},b^\dagger_{0},b^\dagger_{1},b^\dagger_{A}\right) \,,
\end{equation}
where the position in the vector corresponds to the position in the ordered six waveguides. The subscript '$A$' stands for auxiliary and the number stands for the basis state of the qubit.

Using the previous assignments, the matrix associated with post-selected CZ is
\begin{equation}
    \overline{\rm CZ}_{\rm ps} = 
    \frac{1}{\sqrt{3}}
    \begin{pmatrix}
    -1 & \sqrt{2} &  0 & 0 & 0 & 0 \\
    \sqrt{2} & 1 & 0  & 0 & 0 & 0 \\
    0 & 0 & -1 & -\sqrt{2} & 0 & 0 \\
    0 & 0 & \sqrt{2} & -1 & 0 & 0 \\
    0 & 0 & 0 & 0 & -1 & -\sqrt{2} \\
    0 & 0 & 0 & 0 & \sqrt{2} & -1 \\
    \end{pmatrix} \,,
\end{equation}
and the action of $\overline{\rm CZ}_{\rm ps}$ is given by the following recipe
\begin{equation}
    \overline{\rm V}_2 \to \overline{\rm CZ}_{\rm ps} \cdot \overline{\rm V}_2 \,.
\end{equation}
However, the assignment of the qubit states to the waveguides is not symmetric between the first and the second qubit.

It is possible to have a symmetric configuration for the two qubits by changing the order of the assignment and consequently changing the CZ operation.

Let's define a new vector as
\begin{equation}
    {\rm V}_2^{\rm T} \equiv \left(a^\dagger_{A},a^\dagger_{0},a^\dagger_{1},b^\dagger_{A},b^\dagger_{0},b^\dagger_{1}\right) \,.
\end{equation}
This vector is related to the previous one by the following map
\begin{equation}
\begin{split}
        {\rm V}_2 =& {\rm D} \cdot \overline{\rm V}_2 \,,
        \\
        {\rm D} =& 
    \begin{pmatrix}
    1 & 0 &  0 & 0 & 0 & 0 \\
    0 & 1 & 0  & 0 & 0 & 0 \\
    0 & 0 & 1 & 0 & 0 & 0 \\
    0 & 0 & 0 & 0 & 0 & 1 \\
    0 & 0 & 0 & 1 & 0 & 0 \\
    0 & 0 & 0 & 0 & 1 & 0 \\
    \end{pmatrix}\,,
\end{split}
\end{equation}
and the post-selected CZ is achieved in the following way
\begin{equation}
\begin{split}
    {\rm V}_2 &\to {\rm CZ}_{\rm ps} \cdot {\rm V}_2 \,,
    \\
    {\rm CZ}_{\rm ps} &= {\rm D} \cdot \overline{\rm CZ}_{\rm ps} \cdot {\rm D}^{\rm T}
    = \frac{1}{\sqrt{3}}
    \begin{pmatrix}
    -1 & \sqrt{2} &  0 & 0 & 0 & 0 \\
    \sqrt{2} & 1 & 0  & 0 & 0 & 0 \\
    0 & 0 & -1 & 0 & -\sqrt{2} & 0 \\
    0 & 0 & 0 & -1 & 0 & \sqrt{2} \\
    0 & 0 & \sqrt{2} & 0 & -1 & 0 \\
    0 & 0 & 0 & -\sqrt{2} & 0 & -1 \\
    \end{pmatrix}\,,
\end{split}
\end{equation}
where the matrix $D$ is $\mathbf{1}\otimes {\rm SWAP}_1^A$ used in Eq.~\eqref{eq:oSWAP_label}.

It is possible to verify that the generic initial state,
\begin{equation}
\begin{split}
    \vert \Psi_2 \rangle_{\rm in} &= \left( A_0 \,\hat{a}_{0}^\dagger + A_1\, \hat{a}_{1}^\dagger \right)
    \left( B_0\, \hat{b}_{0}^\dagger + B_1\, \hat{b}_{1}^\dagger \right) \vert \Omega \rangle
    \\
    \vert \Psi_2 \rangle_{\rm out} &= \,{\rm CZ}_{\rm ps} \, \vert \psi_0 \rangle
        = -\frac{1}{3}\left(  A_0 B_0 \,\hat{a}_{0}^\dagger \hat{b}_{0}^\dagger + A_0 B_1\,\hat{a}_{1}^\dagger\hat{b}_{0}^\dagger 
        + A_0 B_1\,\hat{a}_{0}^\dagger \hat{b}_{1}^\dagger 
        - A_1 B_1 \,\hat{a}_{1}^\dagger \hat{b}_{1}^\dagger\right) \vert \Omega \rangle +\ldots \\
        &= -\frac{1}{3} \sum_{j_1=0,1} \sum_{j_2=0,1} (-)^{j_1\,j_2}\,A_{j_1} B_{j_2} \,\hat{a}_{j_1}^\dagger \hat{b}_{j_2}^\dagger \,\vert \Omega \rangle
        +\ldots
\end{split}
\end{equation}
where the dots contain all the terms that do not preserve the qubit structure.
Note that if $A_0 = A_1 = B_0 = B_1 = \frac{1}{\sqrt{2}}$ (achieved by H gate applied to both qubits initialized to the zero state), by applying H gate to the second qubit in the output state we get
\begin{equation}
\begin{split}
    \left( \mathbf{1}\otimes {\rm H} \right) \vert \Psi_2 \rangle_{\rm out} &= -\frac{1}{3\sqrt{2}}\left(   \,\hat{a}_{0}^\dagger \hat{b}_{0}^\dagger + \,\hat{a}_{1}^\dagger \hat{b}_{1}^\dagger\right) \vert \Omega \rangle +\ldots 
    \implies
    \frac{1}{\sqrt{2}} \left( \vert \mathsf{00} \rangle + \vert \mathsf{11} \rangle \right)
\end{split}
\end{equation}
which is a Bell state.

Now let's add a third qubit, or equivalently three waveguides. 
As before we arrange the creation operators in a vector
\begin{equation}
    {\rm V}_3^{\rm T} \equiv \left(a^\dagger_{A},a^\dagger_{0},a^\dagger_{1},b^\dagger_{A},b^\dagger_{0},b^\dagger_{1},c^\dagger_{A},c^\dagger_{0},c^\dagger_{1}\right) \,.
\end{equation}
If we apply the post-selected CZ firstly between the first and second qubits and then between the second and third qubits as 
\begin{equation}
\begin{split}
    {\rm V}_3 &\to {\rm CZ}_{\rm ps}^{(2,3)} \cdot {\rm CZ}_{\rm ps}^{(1,2)} \cdot {\rm V}_3 \,,
    \\
    {\rm CZ}_{\rm ps}^{(1,2)} &= 
    \footnotesize
    \begin{pmatrix}
    -\frac{1}{\sqrt{3}} & \sqrt{\frac{2}{3}} &  0 & 0 & 0 & 0 & 0 & 0 & 0 \\
    \sqrt{\frac{2}{3}} & \frac{1}{\sqrt{3}} & 0  & 0 & 0 & 0 & 0 & 0 & 0 \\
    0 & 0 & -\frac{1}{\sqrt{3}} & 0 & -\sqrt{\frac{2}{3}} & 0 & 0 & 0 & 0 \\
    0 & 0 & 0 & -\frac{1}{\sqrt{3}} & 0 & \sqrt{\frac{2}{3}} & 0 & 0 & 0 \\
    0 & 0 & \sqrt{\frac{2}{3}} & 0 & -\frac{1}{\sqrt{3}} & 0 & 0 & 0 & 0 \\
    0 & 0 & 0 & -\sqrt{\frac{2}{3}} & 0 & -\frac{1}{\sqrt{3}} & 0 & 0 & 0 \\
    0 & 0 & 0 & 0 & 0 & 0 & 1 & 0 & 0 \\
    0 & 0 & 0 & 0 & 0 & 0 & 0 & 1 & 0 \\
    0 & 0 & 0 & 0 & 0 & 0 & 0 & 0 & 1
    \end{pmatrix}\,,\\
    {\rm CZ}_{\rm ps}^{(2,3)} &= 
    \footnotesize
    \begin{pmatrix}
    1 & 0 & 0 & 0 & 0 & 0 & 0 & 0 & 0 \\
    0 & 1 & 0 & 0 & 0 & 0 & 0 & 0 & 0 \\
    0 & 0 & 1 & 0 & 0 & 0 & 0 & 0 & 0 \\
    0 & 0 & 0 & -\frac{1}{\sqrt{3}} & \sqrt{\frac{2}{3}} &  0 & 0 & 0 & 0 \\
    0 & 0 & 0 & \sqrt{\frac{2}{3}} & \frac{1}{\sqrt{3}} & 0  & 0 & 0 & 0  \\
    0 & 0 & 0 & 0 & 0 & -\frac{1}{\sqrt{3}} & 0 & -\sqrt{\frac{2}{3}} & 0  \\
    0 & 0 & 0 & 0 & 0 & 0 & -\frac{1}{\sqrt{3}} & 0 & \sqrt{\frac{2}{3}}  \\
    0 & 0 & 0 & 0 & 0 & \sqrt{\frac{2}{3}} & 0 & -\frac{1}{\sqrt{3}} & 0  \\
    0 & 0 & 0 & 0 & 0 & 0 & -\sqrt{\frac{2}{3}} & 0 & -\frac{1}{\sqrt{3}}  
    \end{pmatrix}\,,
\end{split}
\end{equation}
the result is not the expected one.

However, if we truncate the auxiliary waveguides between the two CZ operations and equivalently we apply the following non-unitary operation
\begin{equation}
    \hat{\rm P}_{\rm aux} = 
    \begin{pmatrix}
    0 & 0 & 0 & 0 & 0 & 0 & 0 & 0 & 0 \\
    0 & 1 & 0 & 0 & 0 & 0 & 0 & 0 & 0 \\
    0 & 0 & 1 & 0 & 0 & 0 & 0 & 0 & 0 \\
    0 & 0 & 0 & 0 & 0 &  0 & 0 & 0 & 0 \\
    0 & 0 & 0 & 0 & 1 & 0  & 0 & 0 & 0  \\
    0 & 0 & 0 & 0 & 0 & 1 & 0 & 0 & 0  \\
    0 & 0 & 0 & 0 & 0 & 0 & 0 & 0 & 0  \\
    0 & 0 & 0 & 0 & 0 & 0 & 0 & 1 & 0  \\
    0 & 0 & 0 & 0 & 0 & 0 & 0 & 0 & 1  
    \end{pmatrix} \,.
\end{equation}
This operation projects any state into a state with no photons in the auxiliary waveguides. 

So, we can achieve the desired output state as
\begin{equation}
    {\rm V}_3 \to {\rm CZ}_{\rm ps}^{(2,3)} \cdot \hat{\rm P}_{\rm aux} \cdot {\rm CZ}_{\rm ps}^{(1,2)} \cdot {\rm V}_3 \,.
\end{equation}
It is possible to verify that the generic initial state,
\begin{equation}
\begin{split}
    \vert \Psi_3 \rangle_{\rm out} =& \left( A_0 \,\hat{a}_{0}^\dagger + A_1\, \hat{a}_{1}^\dagger \right)
    \left( B_0\, \hat{b}_{0}^\dagger + B_1\, \hat{b}_{1}^\dagger \right)
    \left( C_0\, \hat{c}_{0}^\dagger + C_1\, \hat{c}_{1}^\dagger \right)\vert \Omega \rangle
    \\
    \vert \Psi_3 \rangle_{\rm out} =& \, {\rm CZ}_{\rm ps}^{(2,3)} \cdot \hat{\rm P}_{\rm aux} \cdot {\rm CZ}_{\rm ps}^{(1,2)} \, \vert \Psi_3 \rangle_{\rm ini} \\
    & = \frac{1}{9}\left(  
    A_0 B_0 C_0 \,\hat{a}_{0}^\dagger \hat{b}_{0}^\dagger \hat{c}_{0}^\dagger 
    + A_0 B_0 C_1\,\hat{a}_{0}^\dagger\hat{b}_{0}^\dagger \hat{c}_{1}^\dagger
    + A_0 B_1 C_0\,\hat{a}_{0}^\dagger\hat{b}_{1}^\dagger \hat{c}_{0}^\dagger 
    - A_0 B_1 C_1\,\hat{a}_{0}^\dagger\hat{b}_{1}^\dagger \hat{c}_{1}^\dagger 
    \right.\\
    &\hspace{1cm}\left.+ A_1 B_0 C_0\,\hat{a}_{1}^\dagger\hat{b}_{0}^\dagger \hat{c}_{0}^\dagger 
    + A_1 B_0 C_1\,\hat{a}_{1}^\dagger\hat{b}_{0}^\dagger \hat{c}_{1}^\dagger 
    - A_1 B_1 C_0\,\hat{a}_{1}^\dagger\hat{b}_{1}^\dagger \hat{c}_{0}^\dagger 
    + A_1 B_1 C_1\,\hat{a}_{1}^\dagger\hat{b}_{1}^\dagger \hat{c}_{1}^\dagger
    \right) \vert \Omega \rangle +\ldots \\
    &= \frac{1}{9} \sum_{j_1=0,1} \sum_{j_2=0,1} \sum_{j_3=0,1} (-)^{j_1\,j_2+j_2\,j_3}\,A_{j_1} B_{j_2} C_{j_3} \,\hat{a}_{j_1}^\dagger \hat{b}_{j_2}^\dagger\hat{c}_{j_3}^\dagger \,\vert \Omega \rangle
        +\ldots
\end{split}
\end{equation}
where again the dots contain all the terms that do not preserve the qubit structure.
Note that if $A_0 = A_1 = B_0 = B_1 = C_0 = C_1 = \frac{1}{\sqrt{2}}$ (achieved by H gate applied to all qubits initialized to the zero state), by applying H gate to the second qubit before the second CZ and H gate to second and third qubits after the second CZ
\begin{equation}
\begin{split}
    \left( \mathbf{1}\otimes \mathbf{1}\otimes {\rm H} \right)\cdot {\rm CZ}_{\rm ps}^{(2,3)} \cdot \left( \mathbf{1}\otimes {\rm H}\otimes \mathbf{1} \right) \cdot \hat{\rm P}_{\rm aux} \cdot {\rm CZ}_{\rm ps}^{(1,2)} \, \vert \psi_0 \rangle &= \frac{1}{9\sqrt{2}}\left(   \,\hat{a}_{0}^\dagger \hat{b}_{0}^\dagger\hat{c}_{0}^\dagger + \,\hat{a}_{1}^\dagger \hat{b}_{1}^\dagger\hat{c}_{1}^\dagger\right) \vert \Omega \rangle +\ldots \\
    & \implies
    \frac{1}{\sqrt{2}} \left( \vert \mathsf{000} \rangle + \vert \mathsf{111} \rangle \right)
\end{split}
\end{equation}
which is the GHZ state for three qubits.
\end{document}